\documentclass[lettersize,journal]{IEEEtran}
\usepackage{amsmath,amsfonts}
\usepackage{algorithmic}
\usepackage{algorithm}
\usepackage{array}
\usepackage[caption=false,font=normalsize,labelfont=sf,textfont=sf]{subfig}
\usepackage{textcomp}
\usepackage{stfloats}
\usepackage{url}
\usepackage{verbatim}
\usepackage{graphicx}
\usepackage{cite}
\usepackage{xcolor}

\begin{document}

\title{Graph Frequency Features of Cancer Gene Co-Expression Networks}

\author{Radwa Adel,~and Ercan Engin Kuruoglu* 
\thanks{The authors are with Tsinghua-Berkeley Shenzhen Institute, Shenzhen International Graduate Schools, Tsinghua University, Shenzhen China. emails:}
\thanks{*Corresponding author: kuruoglu@sz.tsinghua.edu.cn.}
\thanks{Radwa Adel: jiayue20@tsinghua.org.cn.}
\thanks{Manuscript received November xx, 2024; revised XXX xx, 2024.}}

\markboth{ ,~Vol.~XX, No.~X, November~2024}%
{Shell \MakeLowercase{\textit{et al.}}: A Sample Article Using IEEEtran.cls for IEEE Journals}

\IEEEpubid{0000--0000/00\$00.00~\copyright~2024 IEEE}

\maketitle

\begin{abstract}
Complex gene interactions play a significant role in cancer progression, driving cellular behaviors that contribute to tumor growth, invasion, and metastasis. Gene co-expression networks model the functional connectivity between genes under various biological conditions. Understanding the system-level evolution of these networks in cancer is critical for elucidating disease mechanisms and informing the development of targeted therapies. While previous studies have primarily focused on structural differences between cancer and normal cell co-expression networks, this study applies graph frequency analysis to cancer transcriptomic signals defined on gene co-expression networks, highlighting the graph spectral characteristics of cancer systems. Using a range of graph frequency filters, we showed that cancer cells display distinctive patterns in the graph frequency content of their gene transcriptomic signals, effectively distinguishing between cancer types and stages. The transformation of the original gene feature space into the graph spectral space captured more intricate cancer properties, as validated by significantly higher F-statistic scores for graph frequency-filtered gene features compared to those in the original space.
\end{abstract}

\begin{IEEEkeywords}
Graph signal processing, graph frequency filtering, graph Fourier transform, gene co-Expression network, differential co-expression analysis, cancer progression, spectral graph theory. 
\end{IEEEkeywords}

\section{Introduction}
\IEEEPARstart{C}{ancer} is a highly complex disease characterized by its ability to sustain proliferative signaling, resist cell death, evade growth suppressors, enable replicative immortality, induce angiogenesis, and activate invasion and metastasis. These abilities of cancerous cells are mainly derived from mutations in oncogenes, tumor suppressors, and DNA repair genes \cite{HANAHAN2011646}. The disease's inherent complexity lies in its ability to affect virtually any body part and manifest distinct molecular, cellular, and systemic characteristics. A considerable challenge in cancer research is the phenomenon of drug resistance and heterogeneity. Cancer cells can exhibit resistance to therapeutic drugs through various genetic and epigenetic mechanisms. Additionally, substantial variability exists among cancer cells within a tumor (intra-tumor heterogeneity) and across patients (inter-tumor heterogeneity), contributing to unpredictable responses to therapy \cite{holohan2013cancer}\cite{Sun2015}. 

High-throughput RNA sequencing (RNA-seq) has revolutionized the field of gene expression profiling, offering comprehensive characterization of transcriptomes in cancer research\cite{Wang2009}. Differential expression analysis is a commonly used approach for identifying key cancer genes by analyzing changes in gene expression profiles across conditions \cite{Soneson2013}\cite{LONG2023}.

It is increasingly recognized that cancer arises from dysregulation across complex biological systems. System-level approaches have been developed to analyze cancer characteristics by considering integrated cellular systems rather than isolated components\cite{tyson2001network}. Co-expression analysis identifies gene modules that are expressed in harmony. A gene co-expression network is built from high-throughput gene expression data, where nodes represent genes, and edges represent interactions between these genes. Clustering techniques are then employed on a similarity matrix, which captures the network's structure to identify co-expressed gene modules \cite{Soneson2013} \cite{zhang2005general}\cite{horvath2008geometric}. 

After identifying co-expressed genes, differential co-expression analysis examines how the functional interactions between co-expressed genes change dynamically across different biological conditions, like normal versus disease states. This method is crucial in deciphering evolutionary patterns and the emergence of new phenotypes \cite{sandhu2015graph}\cite{tesson2010diffcoex}\cite{pettersen2022csdr}\cite{west2012differential}. However, this approach mainly focuses on differential network topology analysis to capture changes in network typologies, to highlight altered regulatory mechanisms, and to describe system-level characteristics of cancer.

Still, a major challenge remains in developing a unified framework that effectively combines the dynamic change of network structure and the gene expression profiles. Graph Signal Processing (GSP) provides a powerful framework to analyze data defined on graphical structures, integrating graph topology and nodal signals, or graph signals \cite{6494675}. GSP techniques, such as the Graph Fourier Transform (GFT) and graph frequency filtering (GFF), enable unique perspectives in graph signal analysis\cite{RICAUD2019474}.

This approach has been utilized in various applications, from image processing tasks like sharpening to detecting malfunctioning network sensors \cite{8347162}. The frequency content of a graph signal characterizes its spatial variability, with smooth, slowly varying signals having less high-frequency content, while non-smooth, rapidly changing signals have more high-frequency content. Notably, GSP has been leveraged in cancer research, aiding in tasks such as ovarian cancer classification and gene selection for diagnosis \cite{DBLP:journals/corr/SegarraHR14}\cite{WANG2021100662}. Moreover, it was found that brain signals while learning a new task and after becoming familiar with it have distinct low and high-frequency patterns\cite{Huang2016}. Inspired by this, our study seeks to explore the behavior of genes with regulatory relationships in both normal and cancer stages I-IV. The goal is to find whether there exists
\newpage
a characteristic behavior of cancer signals in the frequency domain. 
This study specifically concentrates on frequently mutated genes derived from literature-curated pathways relevant to cancer, particularly the PI3K/AKT and P53 pathways \cite{sanchez2018oncogenic}. Notably, interactions between these pathways are pivotal in regulating cell survival versus death. In normal cellular conditions, p53 transcriptionally activates PTEN, suppressing AKT activity, thereby facilitating p53-mediated apoptosis. A hallmark of many cancers is the dysregulation of these interlinked pathways \cite{singh2002p53}. Given this backdrop, it is insightful to probe the co-expression patterns of these two modules across the spectrum of normal cells to cancer stages I-IV using graph signal processing techniques. Examining the expression profiles of cancer biomarker genes with established functional roles can serve as a robust validation for the capability of graph frequency analysis to discern between normal and cancer samples. This could pave the way for an alternative approach to differential co-expression analysis in cancer research.

Graph frequency analysis is applied to four major cancer types -breast, colon, lung, and stomach, in which the dysregulation of the two pathways is analyzed through normal and the four stages of each cancer type. Focusing on these four cancers allows us to study commonalities and heterogeneity in the dysregulation of the pathways across normal and cancer stages in the frequency domain. In this study, we have two primary objectives, both centered around the potential utility of graph frequency analysis in differential co-expression analysis for cancer. The first is to use GFT to examine the genetic signal spatial variability measured with respect to the co-expression networks across normal and cancer stages I-IV. We aim to uncover distinct frequency behaviors in cancer samples, potentially indicative of specific cancer hallmarks. Secondly, we use frequency-filtered graph analysis in the vertex domain to study genes' contribution to the frequency content.

The paper begins with an overview of our data collection and preprocessing techniques (Section II). We then introduce the gene co-expression network construction using distance correlation (Section III-A). We constructed separate positive and negative gene co-expression networks to represent activating and inhibitory interactions between selected oncogenes and tumor suppressor genes. The GSP notions are then introduced in (Section III-B), including graph signals, GFT, inverse (iGFT), and graph frequency filters. In (Section IV-A) we present the results of the association between the Laplacian eigenvalue distribution and the gene co-expression network connectivity across both normal and cancer stages. We explore the frequency contents of cancer genetic signals in (Section IV-B). The paper concludes by discussing our key findings and their implications in future applications (Section V).
\section{Materials}
\subsection{Data Collection}
\begin{figure}[htbp]
   \centering \includegraphics[width=0.8\linewidth]{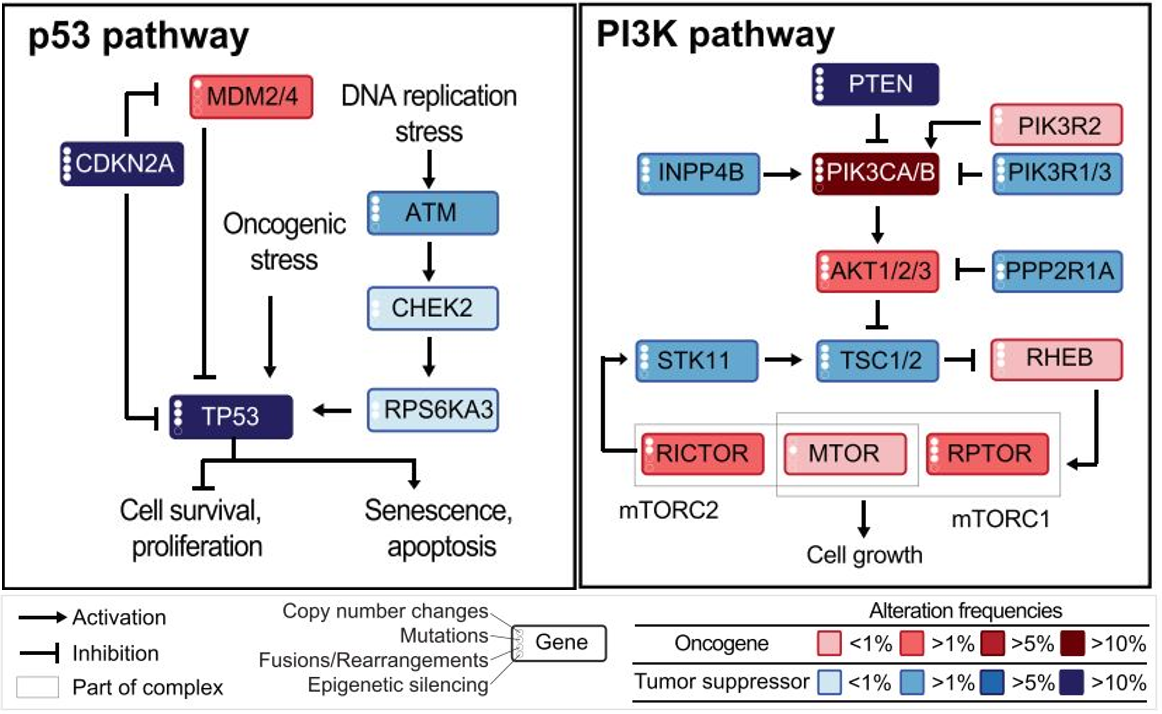} 
   \caption{Curated P53 and PI3K pathways. PI-3-Kinase (PI3K) signaling and P53 pathways members and their interactions. Genes undergo somatic alterations at varying frequencies in cancer samples. The heatmap depicts the average alteration frequency for each gene across the entire dataset, with more intensely colored genes indicating higher average alteration frequencies. Genes altered by oncogenic activations and tumor suppressor inactivation are shown in red and blue, respectively. \cite{sanchez2018oncogenic}.} 
   \label{fig:pathways}
\end{figure}

We use graph spectral analysis on normal and cancer stages I-IV gene expression profiles from breast, colon, lung, and stomach tissues. This enables investigation of cancer progression from healthy to advanced disease stages in the graph frequency domain. We select genes for analysis based on the PI3K signaling and p53 pathways, guided by the study of Sanchez-Vega et al. (2018)\cite{sanchez2018oncogenic}. Our gene selection criteria focus on twenty-three genes highlighted in \cite{sanchez2018oncogenic} for their recurrent genetic alterations and known roles as oncogenes promoting proliferation or tumor suppressors depicted in Figure \ref{fig:pathways}. 

Notably, PIK3R2 and PIK3R3 were excluded due to their minimal expression in normal and cancerous samples. Additionally, while we select genes based on curated biological pathways and known interactions, our analysis mainly utilizes functional connectivity inferred from gene expression profile similarities. In other words, we rely on expression correlation patterns rather than just biological annotations to determine functionally connected genes for our analysis.
The gene expression profiles analyzed were obtained from the TCGA-TARGET-GTEx compendium available on the Xena functional genomics explorer \cite{Goldman326470}, containing samples from The Cancer Genome Atlas (TCGA)\cite{weinstein2013cancer}, Therapeutically Applicable Research to Generate Effective Treatments (TARGET), and Genotype-Tissue Expression (GTEx) \cite{lonsdale2013genotype} projects. These datasets were uniformly reprocessed using the same RNA-seq pipeline to eliminate batch effects [1]. The data consists of RNA-Seq by Expectation-Maximization (RSEM) normalized counts transformed as $\log_{2}(\text{Norm count}+1)$\cite{Goldman326470}. The resulting expression profiles are comparable across projects by reprocessing raw sequences from TCGA, TARGET, and GTEx using a standardized bioinformatics pipeline. Metadata includes sample IDs matching the source studies, primary tissue sites, and sample types (tumor or normal). 

Our study focuses on cancer progression signatures in the graph frequency domain, necessitating cancer stage information to examine disease advancement. Since TCGA provides annotated cancer-stage data, we utilized only TCGA samples for the cancer profiles in our analysis. For normal tissue samples, we leveraged the GTEx project data. By using TCGA and GTEx as the sources of cancer and healthy profiles, respectively, we obtained the gene expression data needed for our graph spectral analysis of cancer progression.
\subsection{Data Pre-Processing}
\begin{figure}[htbp]
	\centering \includegraphics[width=0.8\linewidth]{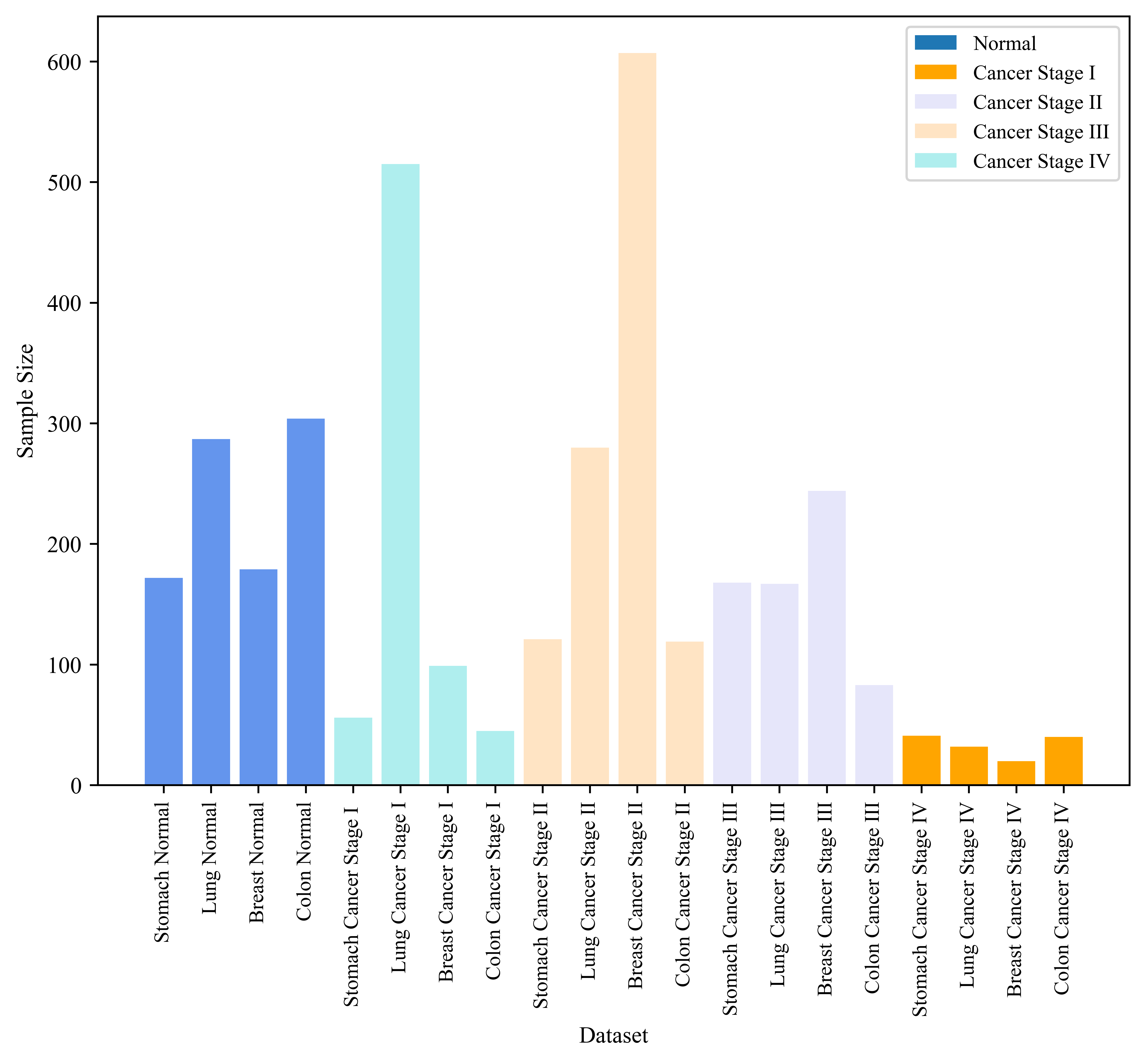} 
	\caption{Normal and Cancer Stages I-TV Dataset Distribution. Each biological condition (normal and cancer stages I-IV) is color-coded across the four types of cancer.} 
	\label{fig:dataset}
\end{figure}

After obtaining the cancer gene expression profiles annotated by TCGA sample IDs, cancer stages I-IV sample ID annotations were collected from the TCGA project  \cite{weinstein2013cancer}. We constructed five gene expression matrices for each cancer type (normal, cancer stage I-. Each data matrix  \( X\) $\in \mathbb{R}^{mxn}$ contains the expression levels of \(n=23\) genes (columns) across \( m\) samples (rows) for the associated biological condition. The total number of samples per condition for each cancer type is shown in Figure \ref{fig:dataset}. We removed significant outlier samples from each data matrix using hierarchical clustering and samples with near zero expressions for all genes for data quality control \cite{langfelder2008wgcna}.
\vspace{-1em}
\subsection{Categorical Trait-Data Construction}
In our study, we aim to discern the associations between graph frequency variables of genetic samples and their respective disease stages, ranging from normal to cancer stages I-IV. Utilizing the established WGCNA \cite{horvath2008geometric} trait categorical data construction method, we constructed a matrix \( T \) $ \in  \mathbb{R}^{qx5}$, where \( q \) signifies the aggregate number of normal and cancer samples. Each column in this matrix corresponds to one of the five categories, with a binary encoding to indicate sample affiliation. Rows represent individual samples through 0/1 vectors indicating their trait classification. We correlate matrix \( T \) with our designated graph spectral variables to uncover variable-trait associations.
\section{Methods}
\subsection{Gene Co-Expression Network Construction}
\subsubsection{Graph Representation}
An undirected weighted graph, \( G \), is used to represent the gene co-expression network and is defined as the pair \( G=(\mathbf{V}, \mathbf{W}) \): 

\begin{itemize}
    \item \textbf{Vertices (V):} Each vertex \( v \in V \), with \( V=\{1,...,n\} \), corresponds to a gene, where \( n \) is the total node count.
    \item \textbf{Weight Matrix (W):} This symmetric matrix \( \mathbf{W}\) $\in \mathbb{R}^{nxn}$ represents the edge weights. Specifically, \( w_{ij} \) indicates the interaction strength between the genes \( i \) and \( j \). The interaction strength is quantified using the distance correlation between their expression profiles \( x_{i} \) and \( x_{j} \) \cite{szekely2009brownian}.
\end{itemize}
\subsubsection{Distance Correlation}
The distance correlation, denoted as \( R(X, Y) \), quantifies the association strength between two random variables and is known to discern both linear and non-linear relations, making it integral for our study \cite{hou2022distance} \cite{szekely2009brownian}.
\begin{equation}
    R(X, Y) = \frac{dCov(X,Y)}{\sqrt{dVar(X)dVar(Y)}}.
\end{equation}
Where:
\begin{equation}
    dCov^{2}(X,Y) = \frac{1}{n^{2}} \sum_{j=1}^{n} \sum_{k=1}^{n} A_{j,k}B_{j,k}.
\end{equation}
The double centered distances matrices \( A_{j,k} \) and \( B_{j,k} \) are formulated as:
\begin{align}
    A_{j,k} &= a_{j,k} - \bar{a_{j.}} - \bar{a_{.k}} + \bar{a_{..}}. \\
    B_{j,k} &= b_{j,k} - \bar{b_{j.}} - \bar{b_{.k}} + \bar{b_{..}}.
\end{align}
Furthermore:
\begin{equation}
    dVar^{2}(X) = \frac{1}{n^{2}} \sum_{j=1}^{n} \sum_{k=1}^{n} A_{j,k}A_{j,k}.
\end{equation}
The key properties of distance correlation are 
\begin{enumerate}
    \item $R(X, Y)$ is defined for X and Y in arbitrary dimensions. 
    \item If $R(X, Y)=0$, $X$ and $Y$ are independent. 
    \item $0\leq R(X,Y)\geq1$.
    \item $R(X,Y)=1$ if and only if $Y=aX+b$, almost surely, with real constants, and $a\neq0$ and $b$.
\end{enumerate}
\subsubsection{Positive and Negative Co-Expression Networks}
To distinguish between positive and negative co-expression networks, Pearson correlation between genes is employed to obtain the sign of the correlations. The weight matrix \( \mathbf{W} \)  is then decomposed into \( \mathbf{W^{+}} \) $in \mathbb{R}^{nxn}$ and \( \mathbf{W^{-}} \) $in \mathbb{R}^{nxn}$, such that $W=W^{+}-W^{-}$ \cite{karaaslanli2022scsgl}. Weights in \( \mathbf{W} \) are conserved only if their associated distance correlation p-values are less than 0.05, ensuring that only statistically significant interactions are considered.
\vspace{-1em}
\subsection{Graph Signal Processing}
This paper focuses on the study of genetic signals defined on co-expression networks. We consider a set of gene expression levels \(x_{i}\) associated with each gene in the network. The set of \(n\) measurements is represented as \(\mathbf{x} = [x_1, x_2, \dots, x_n]^T \in \mathbb{R}^{n}\). A key characteristic of the genetic graph signal \(\mathbf{x}\) is the inherent functional connectivity that defines the values at the vertices of the co-expression network.

In our approach, we utilize GSP concepts rooted in spectral graph theory \cite{8347162}. Specifically, the eigenvalues and eigenvectors of a graph's unnormalized Laplacian matrix correspond to its graph frequencies and graph Fourier basis. We follow conventional definitions of weighted versions of degree and Laplacian matrices as detailed in \cite{chung1997spectral} and \cite{8347162}. The degree matrix $\mathbf{D} \in \mathbb{R}^{n \times n}$ is a diagonal matrix with diagonal elements defined as 
\begin{equation}
\label{wd}
    diag(d_{i}) = \sum_{j \neq i} w_{ij}
\end{equation}
Where $d_{i}$ denotes the weighted degree of the node \(i\). The Laplacian matrix \(\mathbf{L}\) is given by $\mathbf{L} = \mathbf{D} - \mathbf{W}$ with $\mathbf{L} \in \mathbb{R}^{n \times n}$. Being real and symmetric, \(\mathbf{L}\) is positive semi-definite, aligning with the GSP framework requirements derived from spectral graph theory \cite{8347162}.

\subsubsection{Graph Fourier Transform}
Considering the Laplacian matrix \(\mathbf{L}\) is positive semi-definite, it possesses real, non-negative eigenvalues and a complete set of real orthonormal eigenvectors \cite{8347162, ricaud2019fourier}. Applying eigendecomposition to \(\mathbf{L}\) yields:
\begin{equation}
	 \mathbf{L} = \mathbf{U}\Lambda \mathbf{U}^{H}.
\end{equation}
The eigenvalues are organized such that the diagonal matrix \(\Lambda\) has entries: \(\lambda_{i} = \{0 = \lambda_{0} \leq \lambda_{1} \leq \dots \leq \lambda_{n-1}\}\).

The eigenvectors \(\mathbf{U}\) provide an orthonormal basis that represents the network's structure in orthogonal vectors \cite{fiedler1973algebraic, tremblay2014graph, chung1997spectral}. Laplacian eigenvectors tend to be localized, especially in networks with heterogenous degree distributions, implying significant component values for nodes with similar degrees \cite{hata2017localization}.

Given a graph signal \(\mathbf{x}\), Graph Fourier Transform (GFT) maps the signal to the spectral domain using the Laplacian's eigenvectors \cite{8347162}: 
\begin{equation}
\label{GFT}
	\mathbf{U}\tilde{\mathbf{x}} = \mathbf{U}^{H}\mathbf{x}.
\end{equation}
With \(\mathbf{UU^{H}} = \mathbf{I}\), the inverse Graph Fourier Transform (iGFT) of \(\tilde{\mathbf{x}}\) to the vertex domain \(\mathbf{x}\) is:
\begin{equation}
\label{iGFT}
	\mathbf{x} = \mathbf{V}\tilde{\mathbf{x}}.
\end{equation}
The GFT coefficients \(\tilde{\mathbf{x}}\) encodes the magnitude of different spatial variation scales within the graph signal \(\mathbf{x}\) relative to the Laplacian \(\mathbf{L}\). To understand the notion of spatial variability within a graph signal \(\mathbf{x}\), we define the Total Variation (TV) with respect to the Laplacian \(\mathbf{L}\) \cite{ricaud2019fourier} as:
\begin{equation}
\label{tv}
\mathrm{TV}(\mathbf{x}) = \mathbf{x}^{H}\mathbf{L}\mathbf{x} = \sum_{i \neq j} w_{ij} (x_i - x_j)^2.
\end{equation}
From (\ref{tv}), it is evident that the total variation \(\mathrm{TV}(\mathbf{x})\) quantifies the signal's variability with respect to the network topology. Specifically, nodes with higher edge weights \(w_{ij}\) significantly influence the total variation, whereas nodes with smaller \(w_{ij}\) values have a minimal impact on \(\mathrm{TV}(\mathbf{x})\). Now define the total variation of an eigenvector \(\mathbf{TV})\\(\mathbf{u_k})\) as follows:
\begin{equation}
\label{tv_k}
	\mathrm{TV}(\mathbf{u_k}) = \mathbf{u_k^{H}}\mathbf{L}\mathbf{u_k} = \sum_{i \neq j} w_{ij} (u_{ik} - u_{jk})^2 = \lambda_k.
\end{equation}
From (\ref{tv_k}), the eigenvalues \(\lambda_k\) measure the variation of their associated eigenvectors. As the eigenvalues ascend, so does the variation of the eigenvectors. Thus, they offer a notion of frequencies analogous to classical signal processing's eigenmodes. Therefore, when \(k\) is small, the GFT magnitudes $\mathbf{\tilde{x}_{k}}$ of the graph signal indicate the level of spatial variability between weakly connected nodes within the network. On the contrary, as k increases, the GFT magnitudes represent the variability level between closely connected nodes \cite{8347162}.    

\subsubsection{Graph Frequency Filtering}
Graph frequency filtering, as introduced in \cite{huang2016graph}, offers a methodology to isolate specific frequency components, $\mathbf{\tilde{x}_{k}} \in \mathbb{R}^{n \times n}$, of a graph signal. The filtered spectrum is given by:
\begin{equation}
    \mathbf{\tilde{x}_{k}}=\mathbf{H}_{k}\mathbf{\tilde{x}}.
\end{equation}
Where \(\mathbf{H}_{k}\) denotes a diagonal filter matrix, and its diagonal entries, \(\mathbf{h_k}\), are binary such that they are 1 at the desired frequency index \(k\) and 0 elsewhere. The essential goal is to use graph frequency filters to provide insights into the spatial variability of genetic signals across varied connectivity scales inherent to a network's structure. To achieve this, we categorize the frequency spectrum into distinct components: zero-frequency (DC), low, medium, and high. The defined frequency filters are:
\begin{enumerate}
    \item Zero-frequency: $\mathbf{h_0}=\mathbb{I}[k =1]$.
    \item Low-pass: $\mathbf{h_{Lk}}=\mathbb{I}[ k \le  {\lfloor \frac{n-1}{3} \rfloor}]$.
    \item Band-pass: $\mathbf{h_{Mk}}=\mathbb{I}[\lfloor \frac{n-1}{3} \rfloor < k \le  {2\lfloor \frac{n-1}{3} \rfloor}]$.
    \item High-pass: $\mathbf{h_{Hk}}=\mathbb{I}[ k > {2\lfloor \frac{n-1}{3} \rfloor}]$.
\end{enumerate}
These filters allow us to decompose the spectral components of a graph signal into three disjoint frequency components. Because of the eigenvectors' orthogonal nature, they essentially dissect the graph's connectivity and structure \cite{hata2017localization, huang2016graph}. This facilitates the comparative evaluation of graph signal frequencies defined on varying network structures.
We can retrieve filtered signals within the vertex domain by applying the GFT and (iGFT) defined in (\ref{GFT}), (\ref{iGFT}), respectively. The low-pass filtered signal is represented as:
\begin{equation}
\label{f}  
    \mathbf{x_L}=\mathbf{U}\mathbf{\tilde{H}_{L}}\mathbf{U}^{-1}\mathbf{x}.
\end{equation}
Furthermore, the band-pass, high-pass filtered signals can be obtained similarly from (\ref{f}). Therefore, the vertex domain graph signal, \(\mathbf{x}\), is decomposed into three mutually exclusive filtered signals, namely \(\mathbf{x_L}\), \(\mathbf{x_M}\), and \(\mathbf{x_H}\). When the zero-frequency filtered signal is also considered, the original graph signal can be reconstructed:
\begin{equation}
    \label{filter}   
    \mathbf{x}=\mathbf{x_0}+\mathbf{x_L}+\mathbf{x_M}+\mathbf{x_H}.
\end{equation}
Frequency-filtered signals can be understood as \(\mathbf{x_0}\) retains a consistent value across all nodes, equal to the average value across all connected nodes in \(\mathbf{x}\). The variability between each node's value and the average is encompassed within \(\mathbf{x_{Li}}\), \(\mathbf{x_{Mi}}\), and \(\mathbf{x_{Hi}}\). This represents the variance between node \(i\) and its weakly, intermediately, and strongly connected nodes within the network.
\section{Results}
\subsection{Cancer Gene Co-Expression Network Analysis}
This section evaluates the median, variance, and algebraic connectivity (second smallest eigenvalue) of positive and negative gene co-expression networks for normal and cancer stages I-IV. Disconnected small clusters or nodes were removed from the weight matrix, ensuring consistent graph frequency analysis across different network structures. Our primary objective is to discern global characteristics of the dynamic changes in activating (positive) and inhibitory (negative) interactions as cancer progresses.
\subsubsection{Analysis of Eigenvalue and Weighted Degree Distributions in Co-expression Networks}
We compared eigenvalue distribution metrics to the weighted degree defined in (\ref{wd}) distribution of the 40 gene co-expression networks to validate the association between network connectivity and Laplacian eigenvalues. The median eigenvalue exhibited a substantial correlation ($r=0.8, p<0.05$) with the average weighted degree. The second smallest eigenvalue correlated ($r=0.6, p<0.05$) with network density as defined in \cite{horvath2008geometric}, which measures how tightly the network is connected. Notably, the variance of the eigenvalue distribution significantly correlated ($r=0.9, p<0.05$) with the variance of the weighted degrees. Therefore, the Laplacian eigenvalues strongly reflect the connectivity pattern of gene co-expression networks.
\subsubsection{Comparative Analysis of Eigenvalue Distribution in Normal and Cancer Co-Expression Networks}
\begin{figure}[!t]
	\centering
	\subfloat[\textrm{\footnotesize{Breast Positive Network}}]{\includegraphics[width=1.5in]{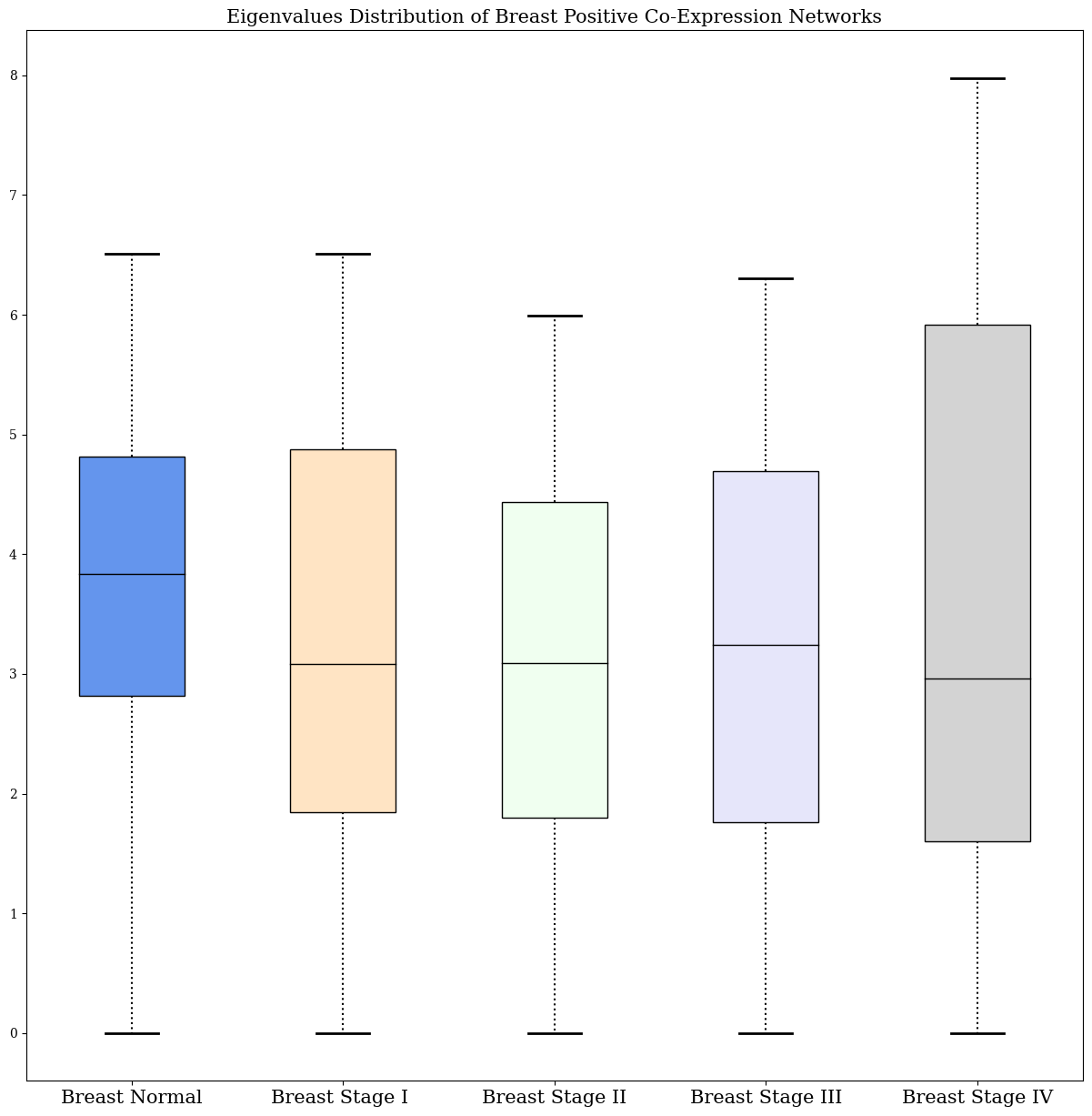}%
		\label{breast_net_p}}
	\subfloat[\textrm{\footnotesize{Breast Negative Network}}]{\includegraphics[width=1.5in]{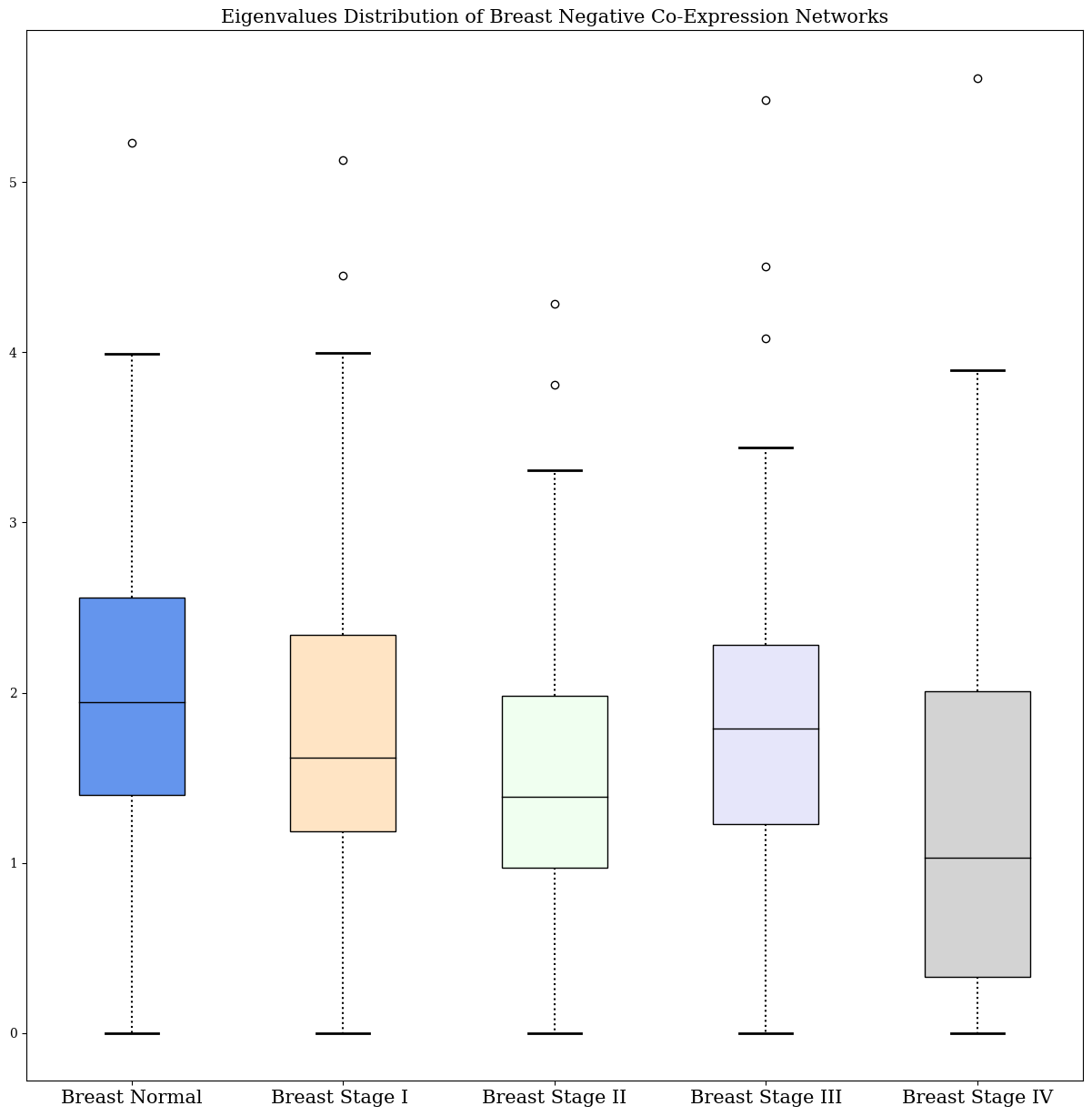}%
		\label{breast_net_n}}

	\subfloat[\textrm{\footnotesize{Colon Positive Network}}]{\includegraphics[width=1.5in]{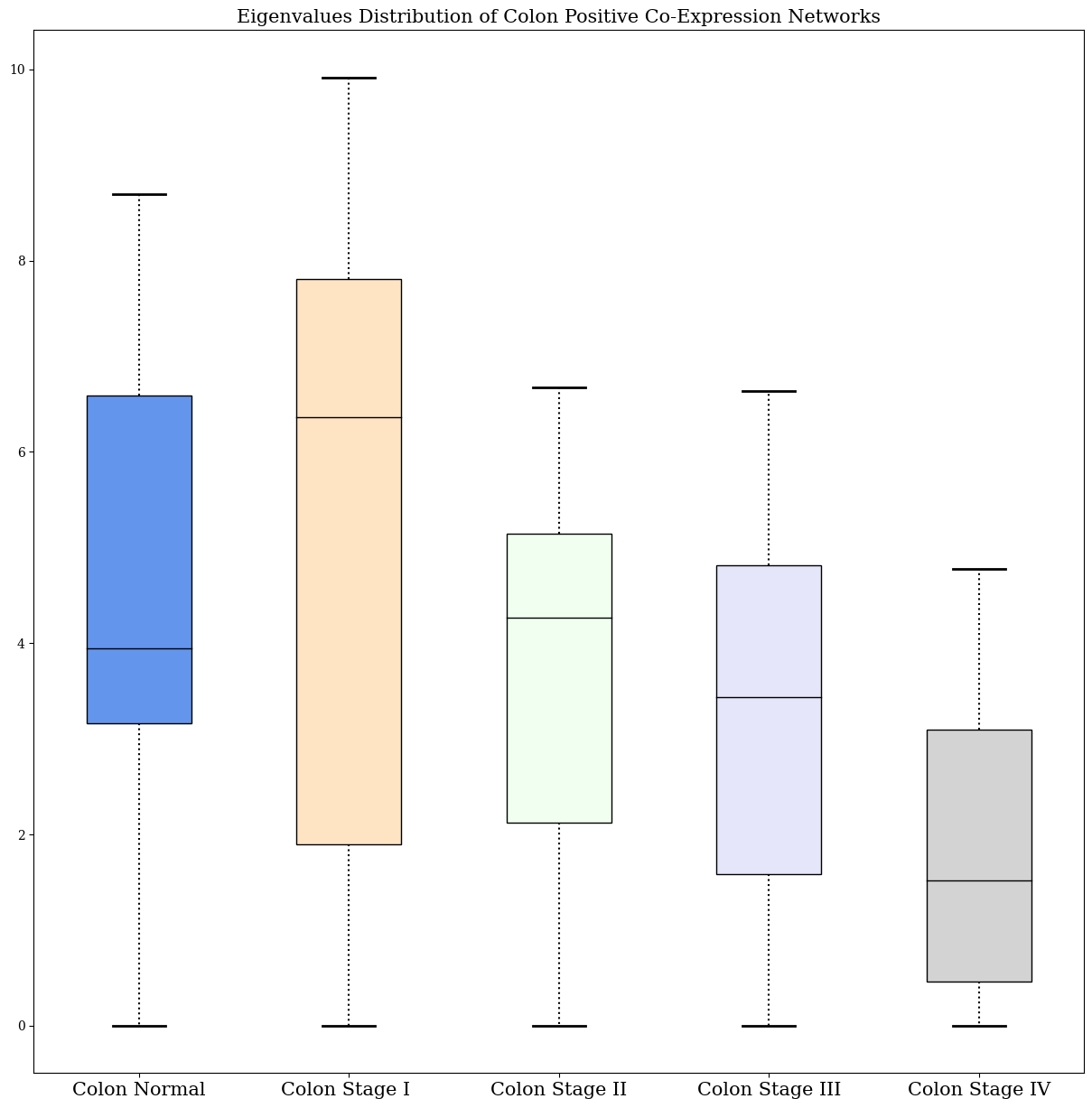}%
		\label{colon_net_p}}
	\subfloat[\textrm{\footnotesize{Colon Negative Network}}]{\includegraphics[width=1.5in]{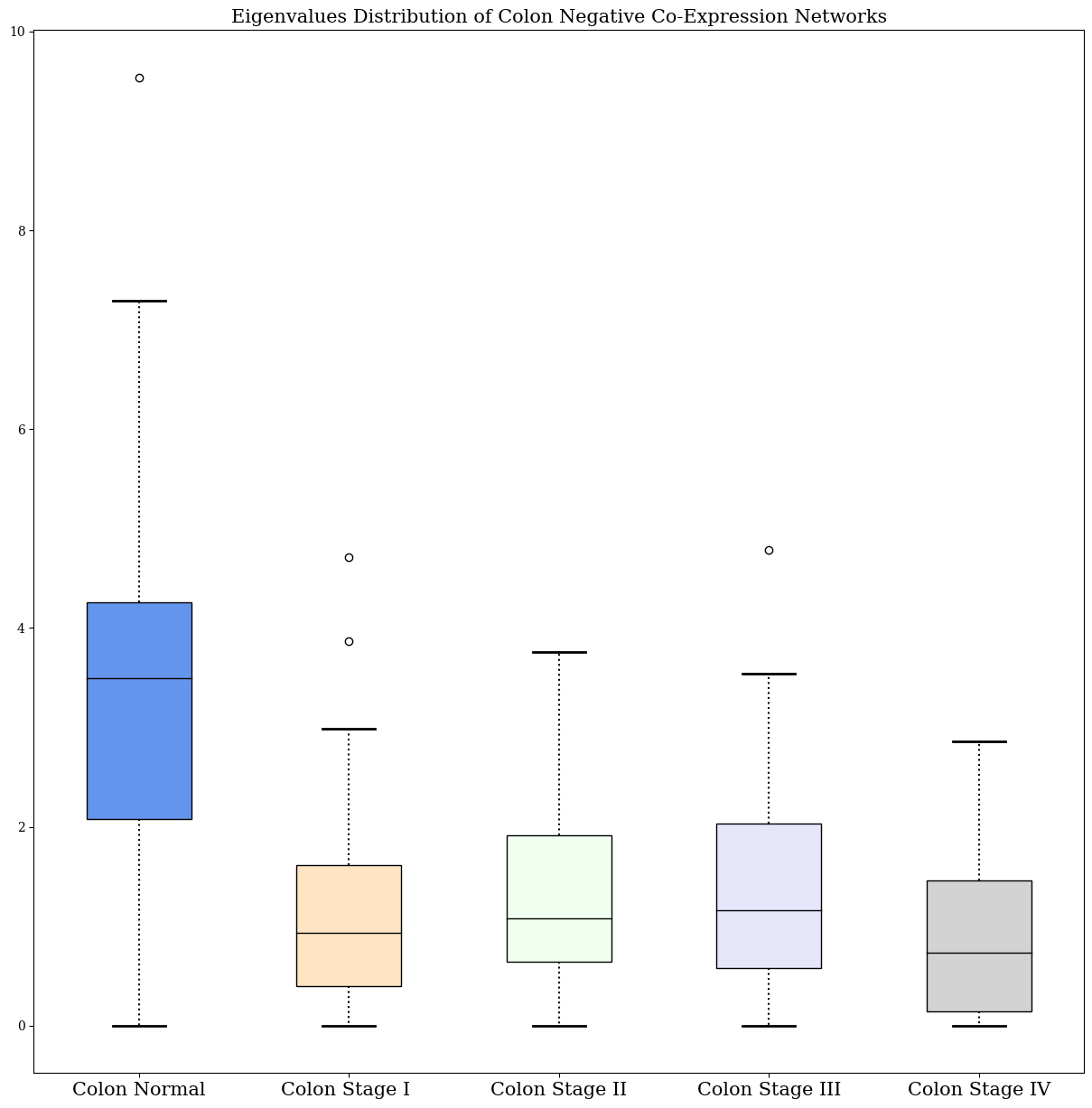}%
		\label{colon_net_n}}

	\subfloat[\textrm{\footnotesize{Lung Positive Network}}]{\includegraphics[width=1.5in]{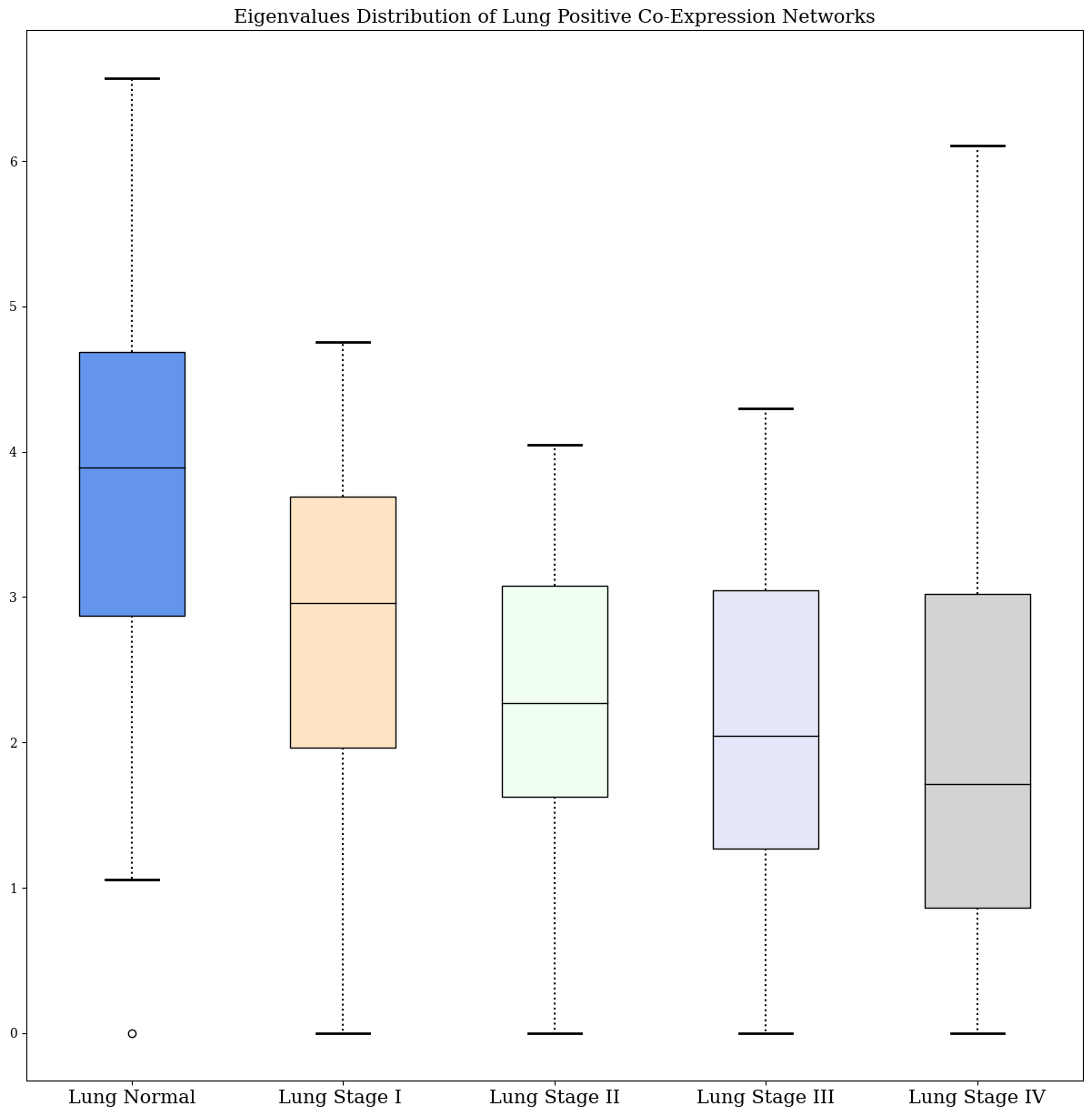}%
		\label{lung_net_p}}
	\subfloat[\textrm{\footnotesize{Lung Negative Network}}]{\includegraphics[width=1.5in]{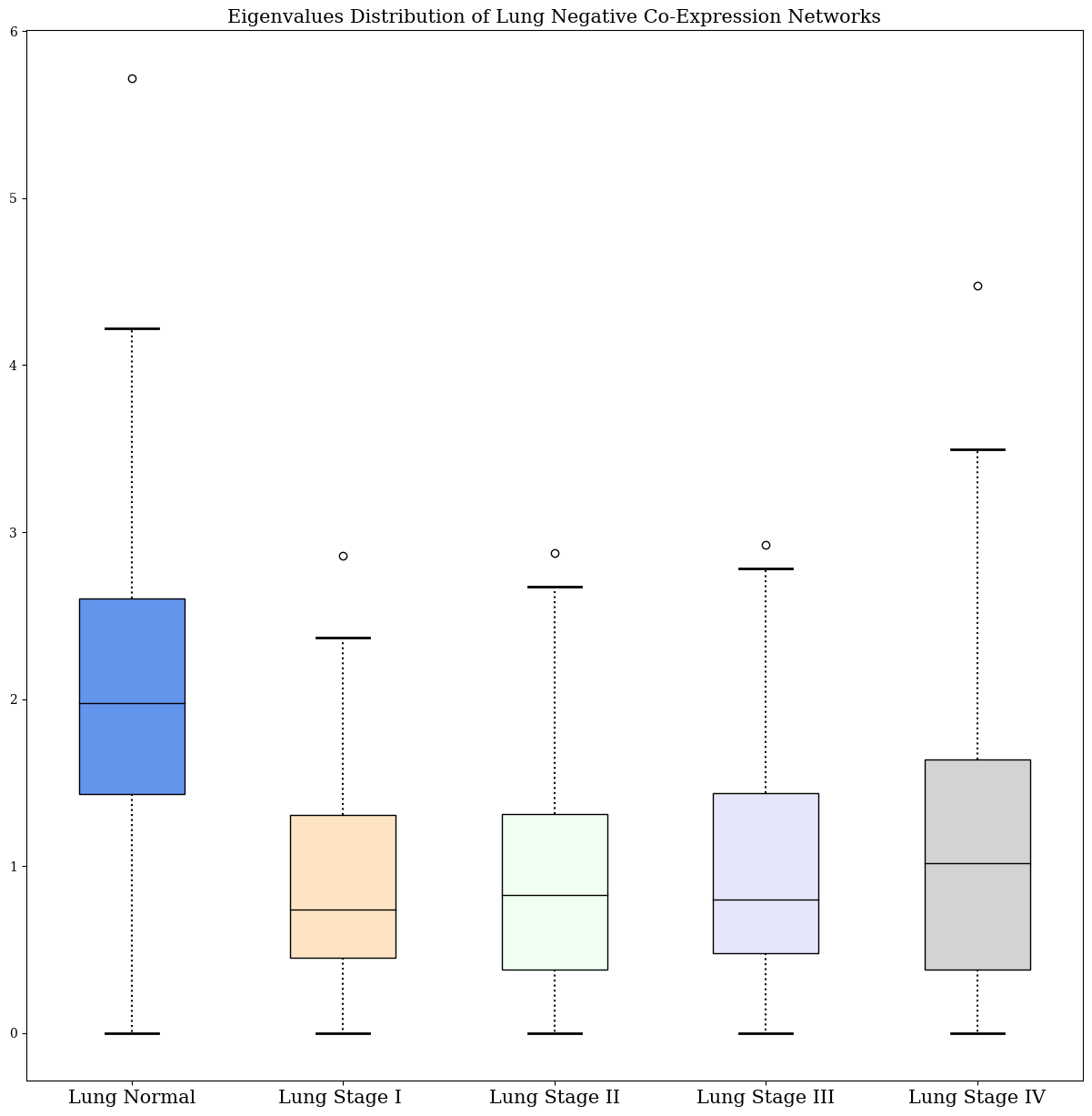}%
		\label{lung_net_n}}

	\subfloat[\textrm{\footnotesize{Stomach Positive Network}}]{\includegraphics[width=1.5in]{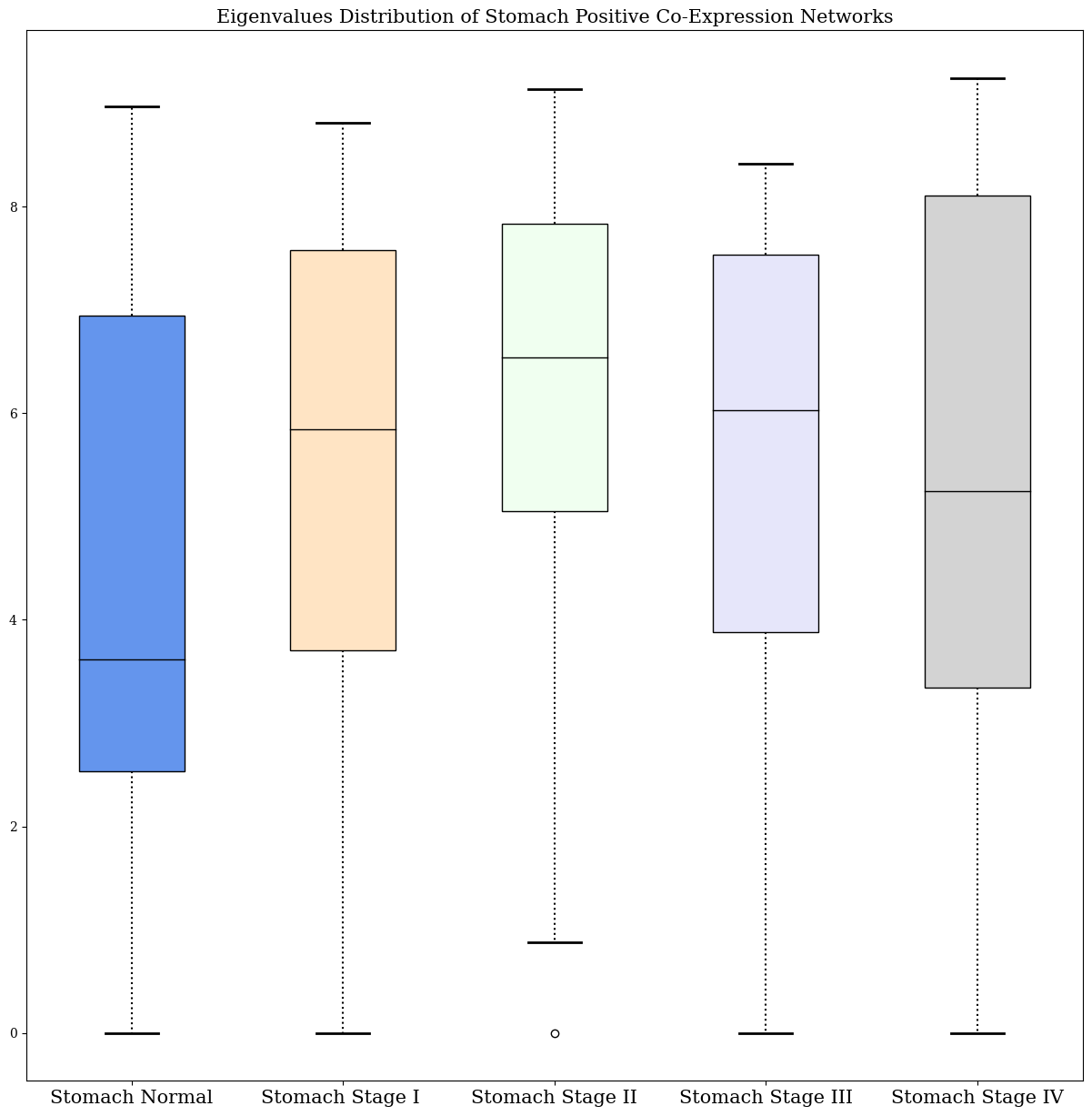}%
		\label{stomach_net_p}}
	\subfloat[\textrm{\footnotesize{Stomach Negative Network}}]{\includegraphics[width=1.5in]{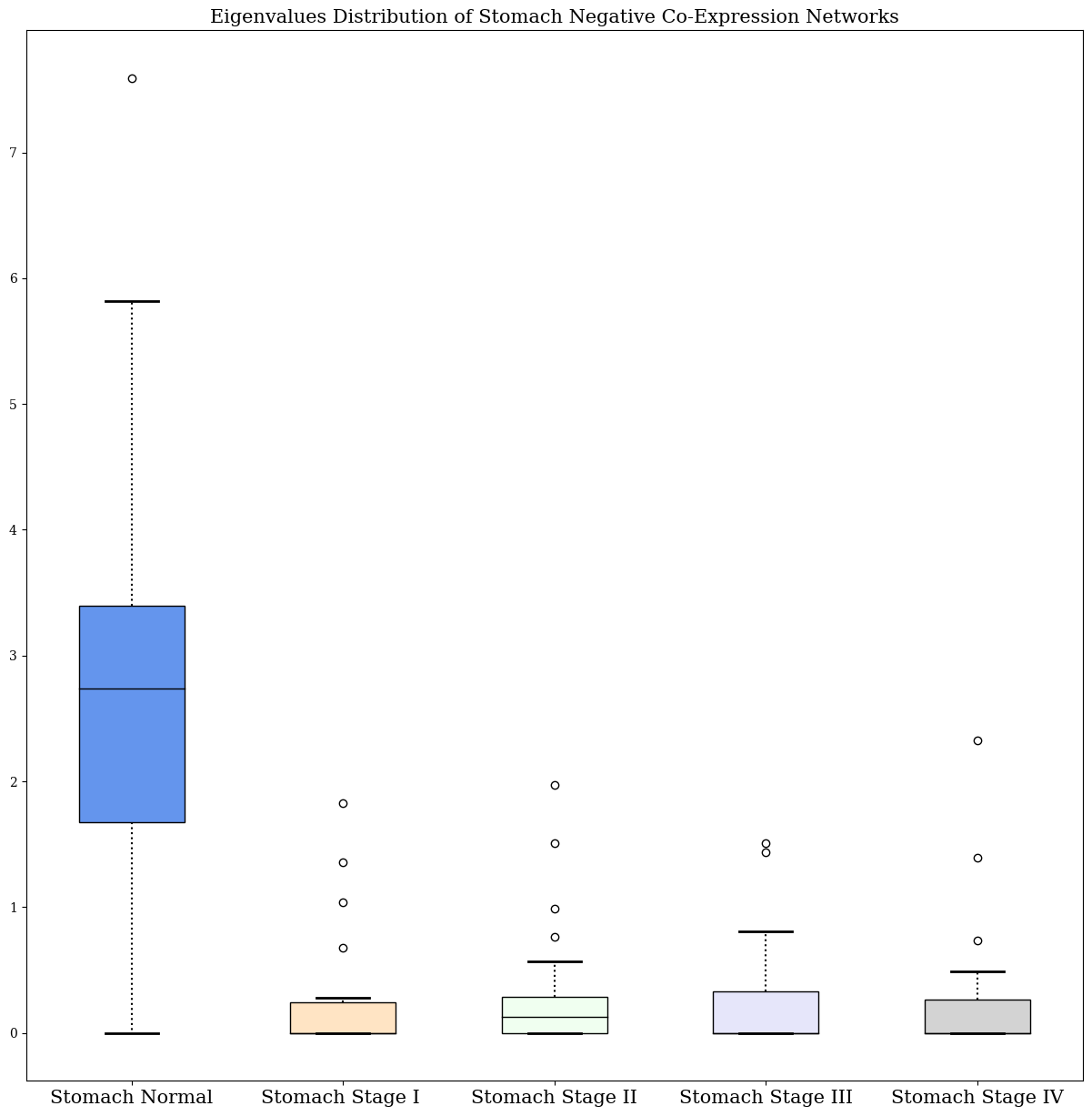}%
		\label{stomach_net_n}}
	\caption{Eigenvalue Distribution of Positive and Negative Gene Co-expression Network Laplacians.  Box plots of eigenvalue distributions for gene co-expression network Laplacians across normal and cancer progression stages I-IV. Positive and negative Laplacian eigenvalue distributions are shown for four tissue types: (a, b) breast, (c, d) colon, (e, f) lung, and (g, h) stomach. The eigenvalue distributions are given for each tissue type, from normal tissue (leftmost, blue) to cancer stage IV (rightmost, grey).}
	\label{eig}
\end{figure}

In \textbf{breast} cancer co-expression networks, both \textbf{positive} and \textbf{negative}, we observed a declining trend in the Laplacian eigenvalues' median and algebraic connectivity with cancer progression. This implies a global \textbf{reduction} in the strength of positive and negative interactions. Conversely, eigenvalue variance exhibited a high increase in the positive networks.  Therefore, this indicates that some genes become more connected than others, as depicted in Figures \ref{breast_net_p} and \ref{breast_net_n}. 
 Additionally, within \textbf{lung} cancer positive and negative co-expression networks, a trend of decreased connectivity analog to those in \textbf{breast} cancer was observed as shown in Figures \ref{lung_net_p} and \ref{lung_net_n}. However, the decrease in the median of the eigenvalues was more significant than in breast cancer. 

We observed a marked \textbf{rise} in the eigenvalues' median and variance in \textbf{colon} cancer stage I \textbf{positive} co-expression network, with a minor increase in algebraic connectivity, as demonstrated in Figure \ref{colon_net_p}. However, these values diminished with further cancer progression. Intriguingly, Figure \ref{colon_net_n} shows that the \textbf{negative} co-expression networks displayed a significant \textbf{decrease} in the eigenvalues' median, variance, and algebraic connectivity. This suggests stronger positive interactions as cancer initiates and weaker negative interactions between the genes in colon cancer. However, because of the high variance in the eigenvalues, and most eigenvalues are below the median, this indicates the existence of genes significantly more connected than others.

Interestingly, in the \textbf{positive} co-expression networks of \textbf{stomach} cancer stages I-IV, there was a \textbf{significant rise} in both the median of the eigenvalues and the algebraic connectivity relative to the normal network, paired with a slight drop in variance. Conversely, the \textbf{negative} co-expression networks experienced a \textbf{notable reduction} in the strength of the negative interactions across all cancer stages, as illustrated in Figures \ref{stomach_net_p} and \ref{colon_net_p}.
\vspace{-1em}
\subsection{Graph Frequency Analysis of Cancer Genetic Signals}
\subsubsection{Graph Frequency Analysis in the Spectral Domain}
Our previous analysis revealed an association between Laplacian eigenvalues and weighted gene degrees, aligning with findings in \cite{hata2017localization}. This motivated analyzing cancer genetic signals (gene expression levels) in the frequency/spectral domain through the GFT defined in (\ref{GFT}). We examined the change of the zero $\left\| \tilde{x}_{0} \right\|_2$, low-frequency $\left\| \tilde{x}_{L} \right\|_2$, medium-frequency $\left\| \tilde{x}_{M} \right\|_2$, and high-frequency $\left\| \tilde{x}_{H} \right\|_2$ components (magnitudes) within the genetic signal onto positive and negative co-expression networks. 

The key difference in the frequency components of the genetic signal between positive and negative networks lies in the expected dominant components. In positive networks, lower frequencies are expected to have higher magnitudes, as spatial variability arises primarily from weakly connected genes by definition. In contrast, high frequencies are expected to dominate within negative networks, as closely connected genes contribute most to spatial variability due to their strong inhibitory associations.

Positive networks' DC component (zero-frequency magnitude) indicates average expression across interconnected genes. Heightened DC, given constant connections, implies over-expression between conditions. In negative networks, a diminished DC magnitude accompanied by higher magnitudes of other frequency ranges suggests increased co-existence of under-expressed and over-expressed genes, indicating stronger underlying inhibitory regulation.

\begin{figure*}[htbp]
    \centering
    \begin{minipage}{\linewidth}
        \centering
        \subfloat[\textrm{\footnotesize{DC-Trait (Positive)}}]{\includegraphics[width=1.5in]{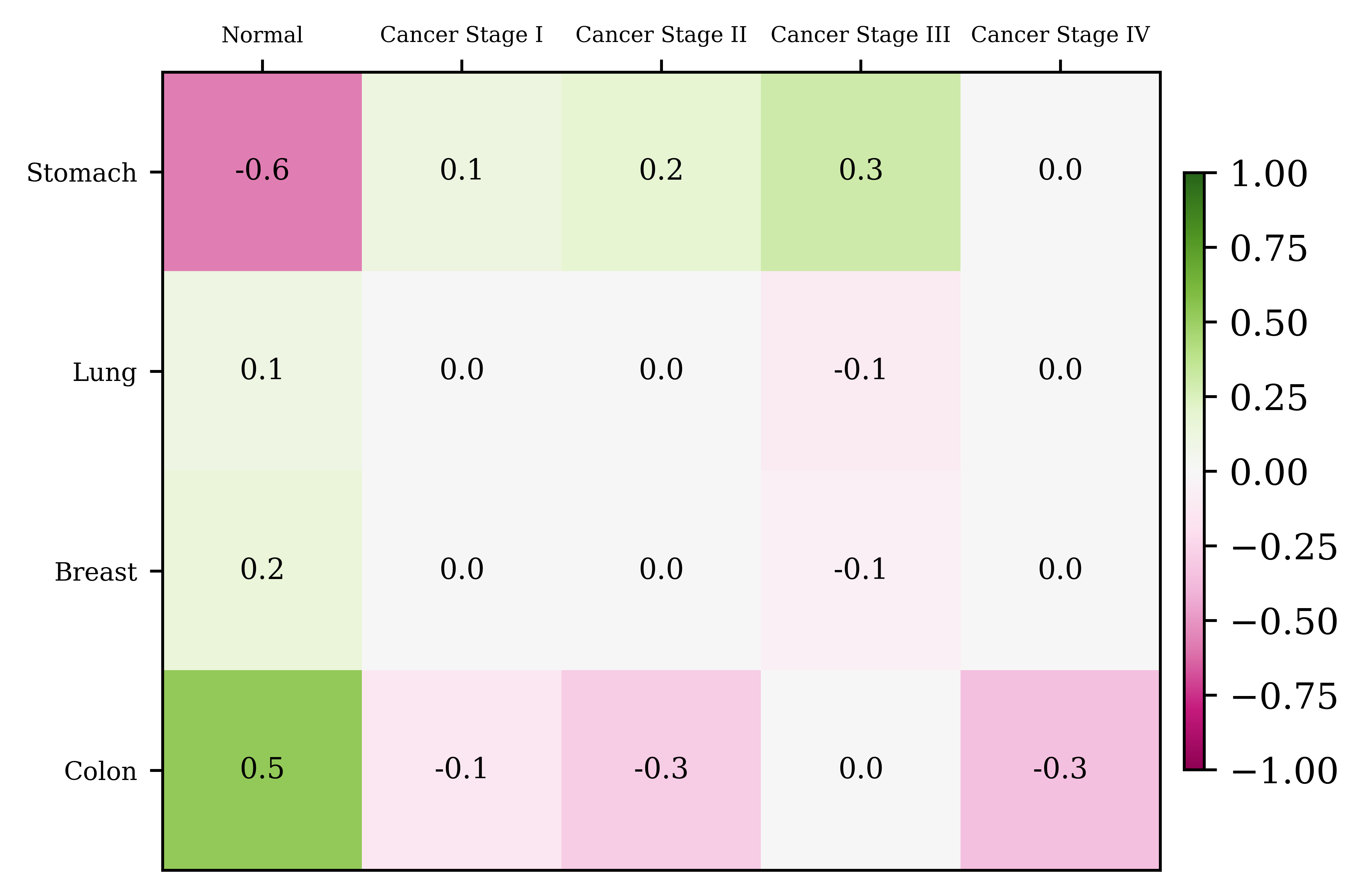}\label{fig:DC_p}}\hfil
        \subfloat[\textrm{\footnotesize{Low Freq-Trait (Positive)}}]{\includegraphics[width=1.5in]{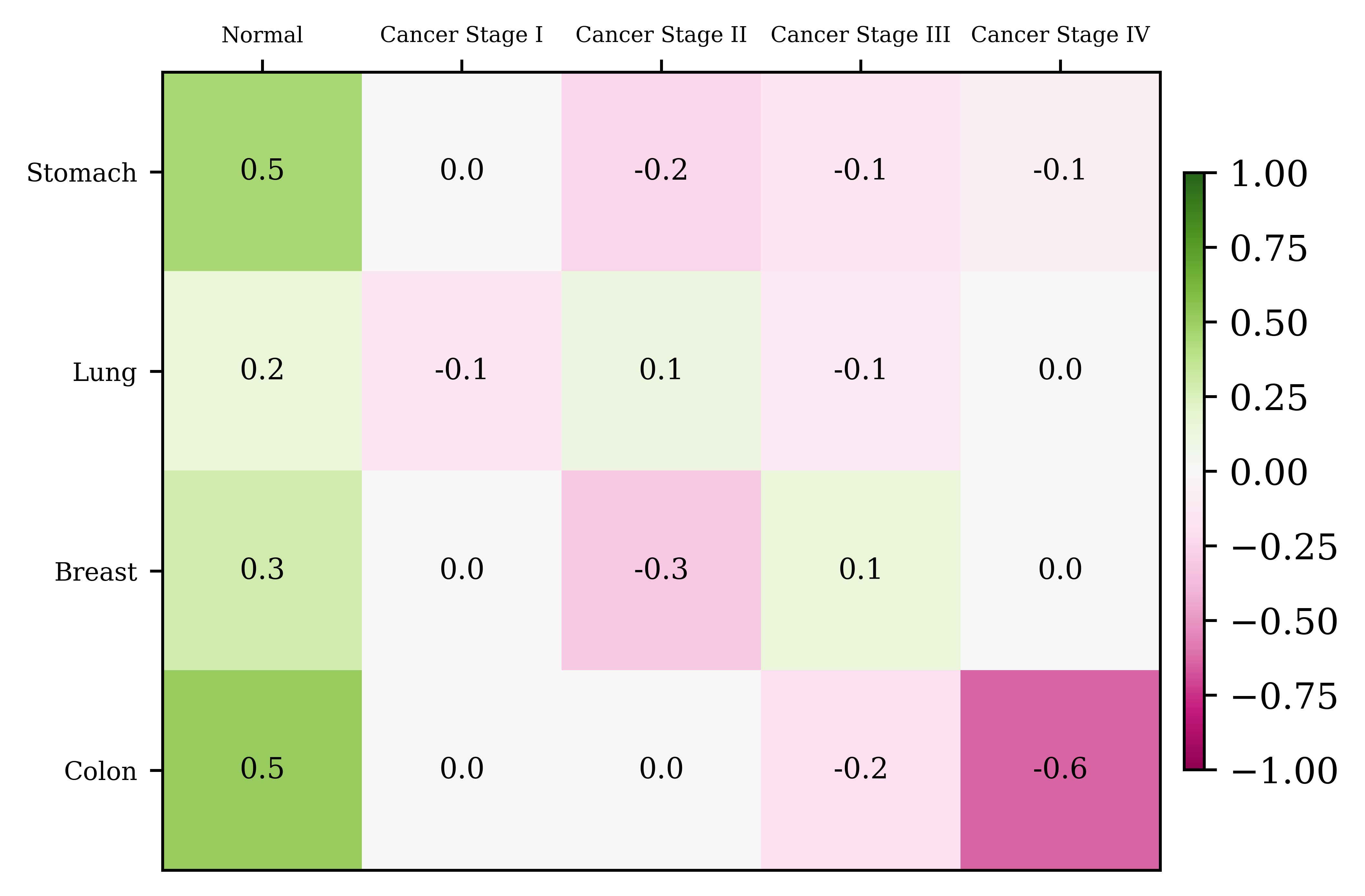}\label{fig:low_p}}\hfil
        \subfloat[\textrm{\footnotesize{Medium Freq-Trait (Positive)}}]{\includegraphics[width=1.5in]{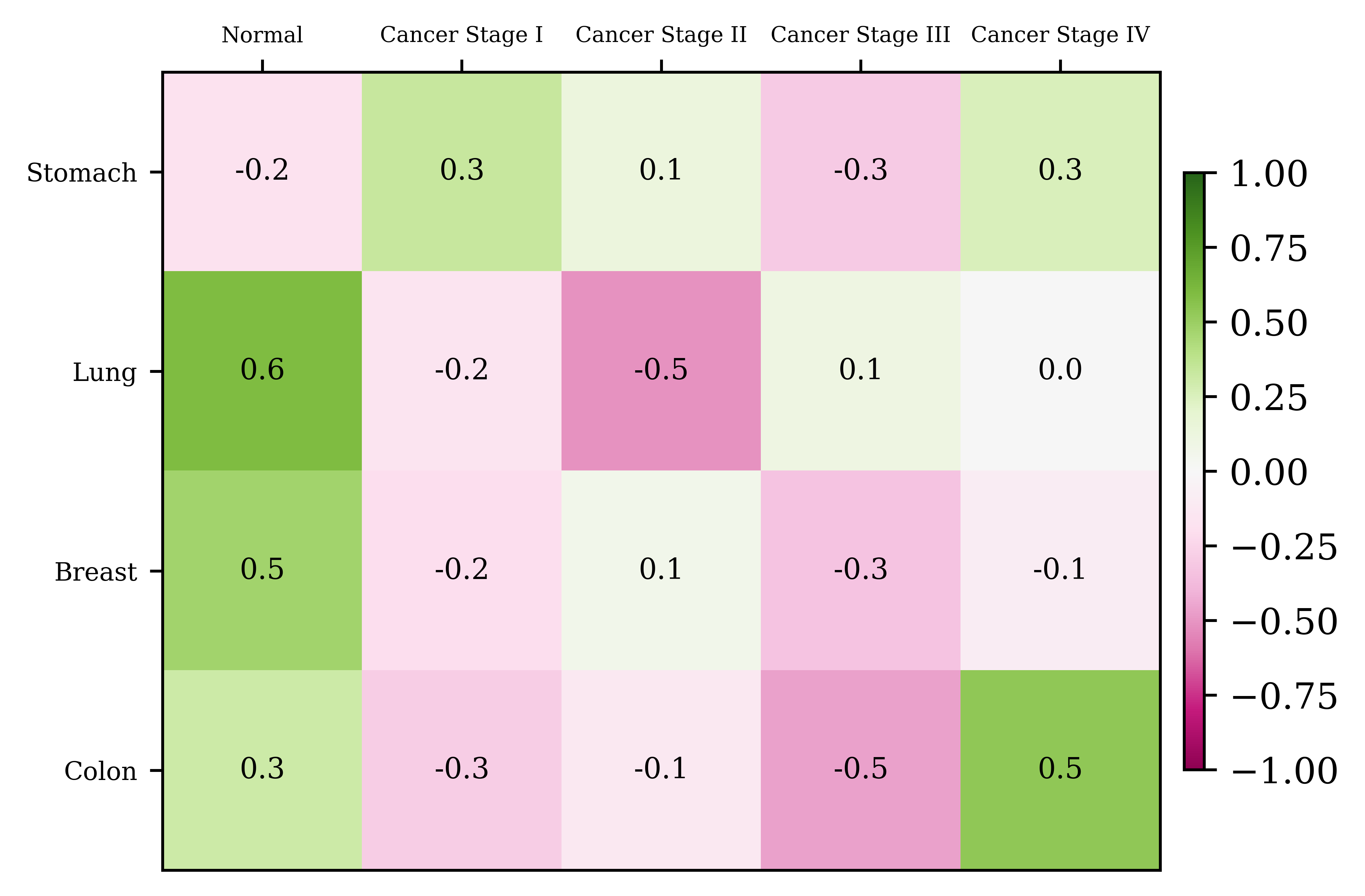}\label{fig:medium_p}}\hfil
        \subfloat[\textrm{\footnotesize{High Freq-Trait (Positive)}}]{\includegraphics[width=1.5in]{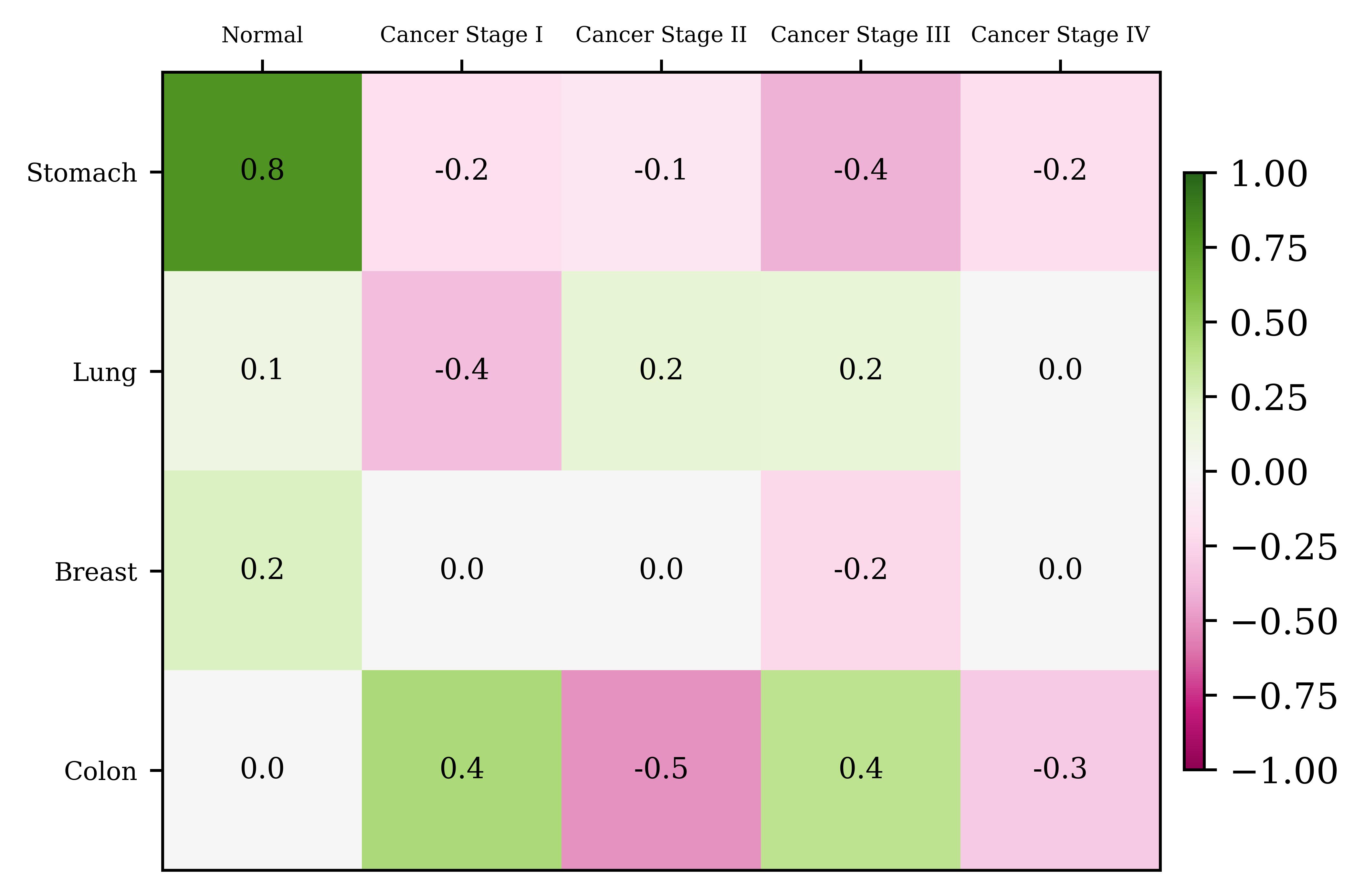}\label{fig:high_p}}
    \end{minipage}

    \vspace{0.5cm}

    \begin{minipage}{\linewidth}
        \centering
        \subfloat[\textrm{\footnotesize{DC-Trait (Negative)}}]{\includegraphics[width=1.5in]{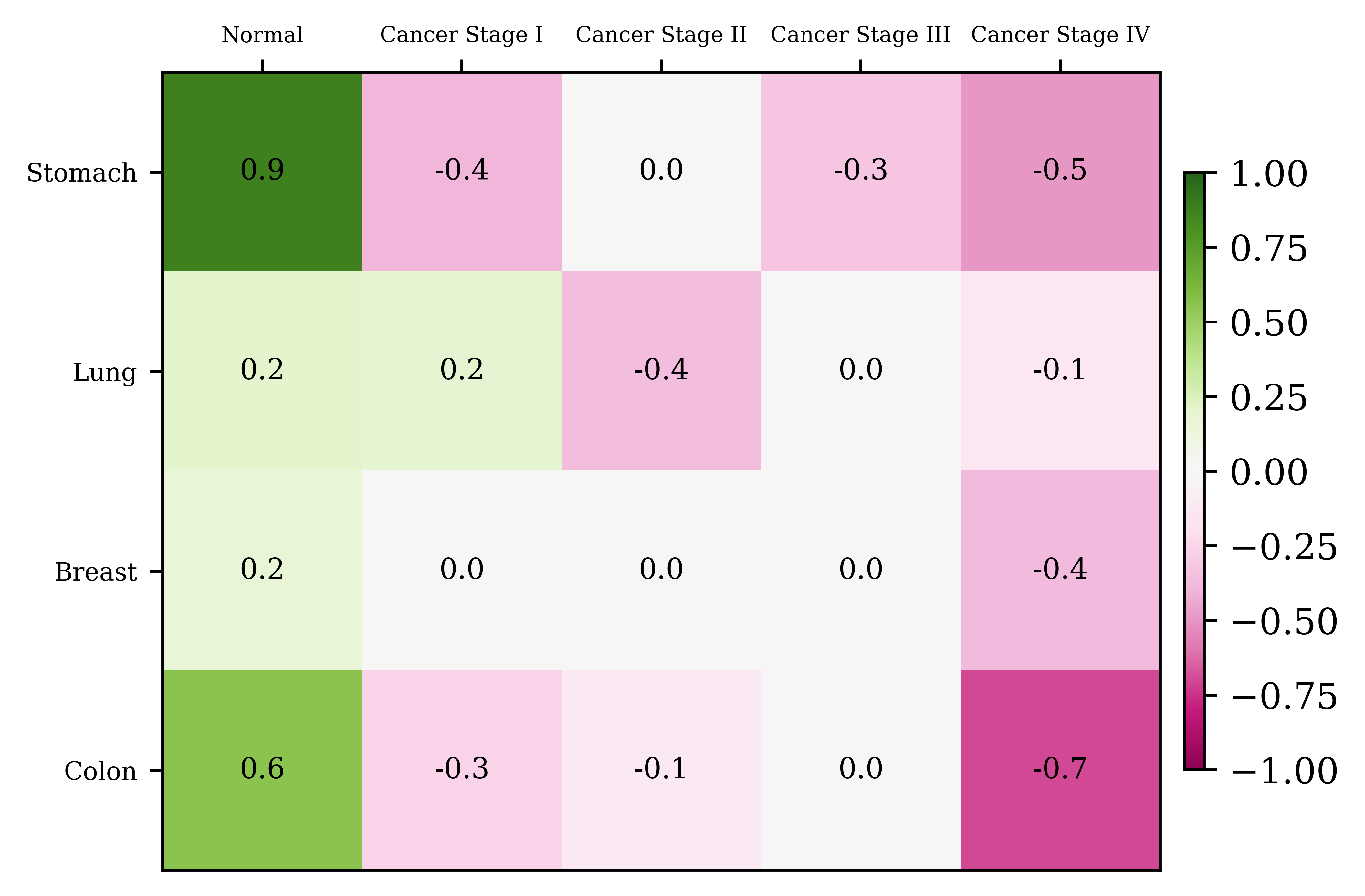}\label{fig:DC_n}}\hfil
        \subfloat[\textrm{\footnotesize{Low Freq-Trait (Negative)}}]{\includegraphics[width=1.5in]{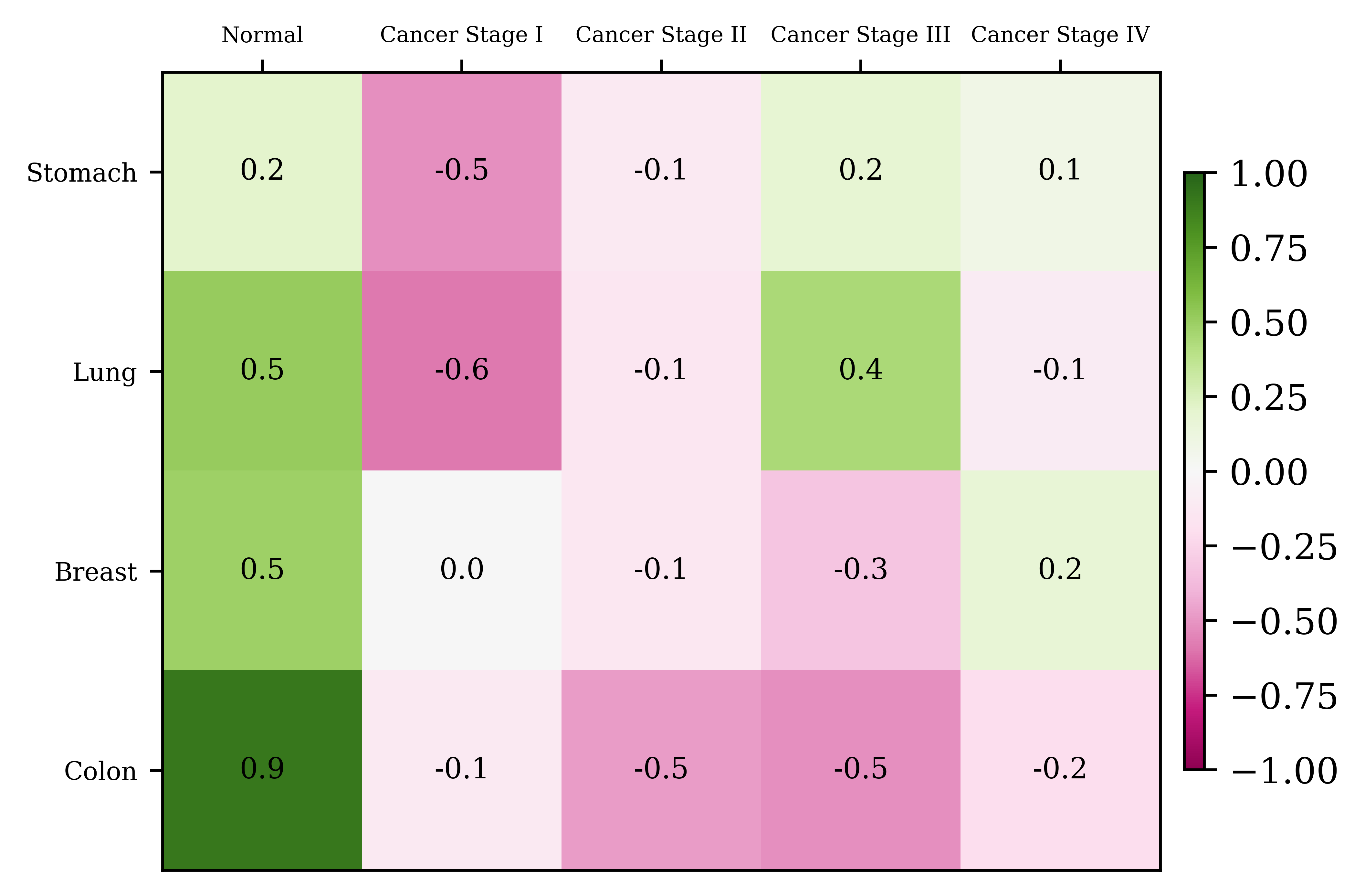}\label{fig:low_n}}\hfil
        \subfloat[\textrm{\footnotesize{Medium Freq-Trait (Negative)}}]{\includegraphics[width=1.5in]{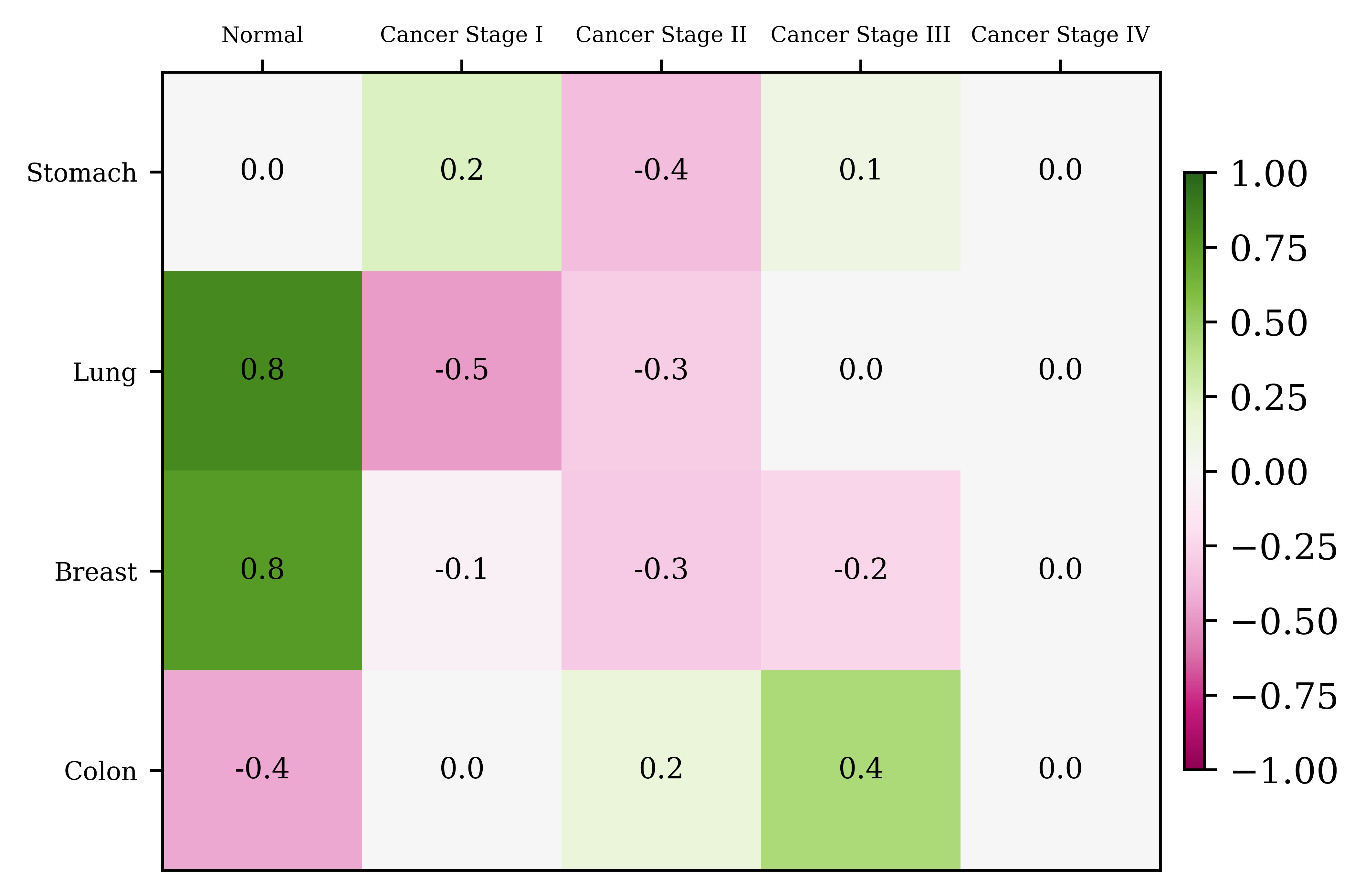}\label{fig:medium_n}}\hfil
        \subfloat[\textrm{\footnotesize{High Freq-Trait (Negative)}}]{\includegraphics[width=1.5in]{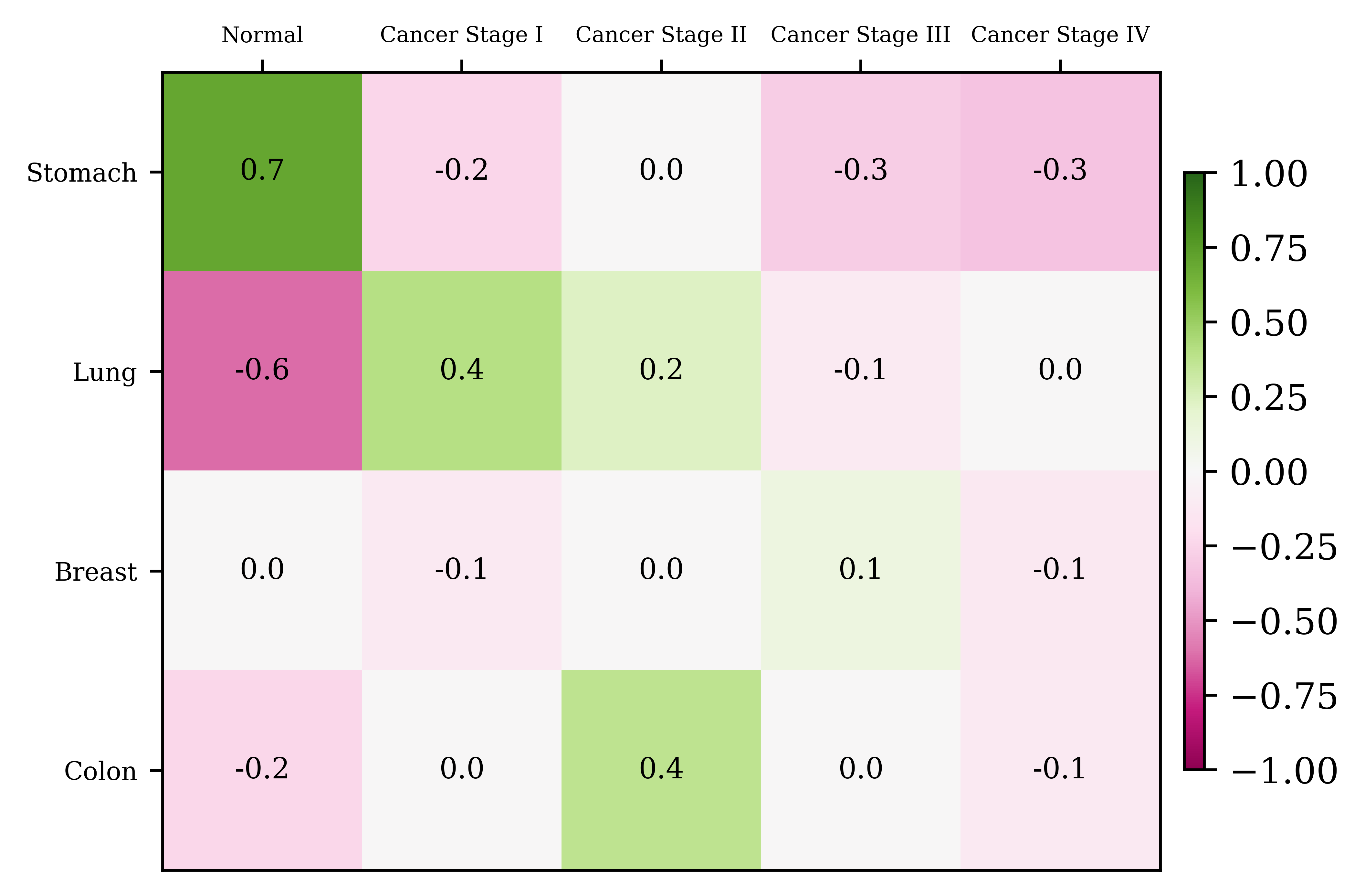}\label{fig:high_n}}
    \end{minipage}

    \caption{Association Between Genetic Signals Frequency Components With The Biological Traits of Normal and Cancer Stages (I-IV). This figure presents heatmaps of Pearson correlation coefficients, illustrating the relationship between the amplitudes of Direct Current (DC), low, medium, and high frequencies of genetic signals when projected onto the eigenvectors of positive (a-d) and negative (e-h) networks of the four types of cancer. Each heatmap is organized such that the rows represent tissue type (stomach, lung, breast, and colon), and the columns represent the biological condition. All presented correlations have p-values smaller than 0.05.}
    \label{GFT_combined}
\end{figure*}

To determine statistically significant associations between the frequency components of the genetic signals and the stages of cancer, Pearson correlations were calculated between trait variables $\mathbf{T_i}$, defined earlier, and the amplitudes of the frequency components $\text{cor}(\left\| \tilde{x}_{k} \right\|_2, \mathbf{T_i})$. Figure \ref{GFT_combined} presents the correlation results for positive and negative network frequencies. All highlighted correlations have $p < 0.05$. The key findings of the change of the spectral components in normal and cancer samples on positive and negative networks are presented in Tables \ref{tab:spectral_behavior} and \ref{tab:spectral_behavior_negative}, respectively. 

\begin{table*}[htbp]
\centering
\caption{Spectral Behavior of Cancer Genetic Signals on Positive Interactions}
\label{tab:spectral_behavior}
\begin{tabular}{|c||p{13cm}|}
\hline
\textbf{Key Findings} & \textbf{Description} \\
\hline
General Trend & Across all cancer types, genetic signals showed \textbf{decreased low}-frequency magnitude (Figure \ref{fig:low_p}). \\
\hline
Breast and lung cancer & Early-stage breast and lung cancers displayed a notable \textbf{decrease for medium and high}-frequency magnitude (Figures \ref{fig:medium_p} and \ref{fig:high_p}).\\
\hline
Stomach Cancer & Signals exhibited \textbf{higher DC} magnitudes and \textbf{diminished high}-frequency magnitudes compared to normal tissue (Figure \ref{fig:DC_p} and \ref{fig:high_p}). \\
\hline
Colon Cancer & A marked \textbf{reduction in DC} magnitude was observed in all cancer stages. In stage I, an \textbf{increase in the high}-frequency component magnitude was found. This pattern becomes inconsistent in advanced stages (Figure \ref{fig:DC_p} and \ref{fig:high_p}). \\
\hline
\end{tabular}
\end{table*}
\begin{table*}[htbp]
\centering
\caption{Spectral Behavior of Cancer Genetic Signals on Negative Interactions}
\label{tab:spectral_behavior_negative}
\begin{tabular}{|c||p{13cm}|}
\hline
\textbf{Key Findings} & \textbf{Description} \\
\hline
General Trends & All cancer types showed a notable \textbf{reduction in low}-frequency magnitudes (Figure \ref{fig:low_n}). \\
\hline
Breast and lung cancer & Signals exhibited a distinct \textbf{drop in the medium}-frequency magnitude (Figure \ref{fig:medium_n}).\\
\hline
Lung Cancer &  Signals on negative interactions revealed a \textbf{rise in high}-frequency magnitudes (Figure \ref{fig:high_n}). \\
\hline
Stomach Cancer & The signals displayed a marked \textbf{decrease in DC} magnitudes (Figure \ref{fig:DC_n}), pointing to the disconnection in the negative cancer networks. Signals also exhibited \textbf{drop in high}-frequency magnitudes, indicating smoother expression signals (Figure \ref{fig:high_n}). \\
\hline
Colon Cancer & A noticeable \textbf{increase in the medium}-frequency magnitude (Figure \ref{fig:medium_n}), suggesting heightened variability in expression levels for genes with intermediate connectivity. \\
\hline
\end{tabular}
\end{table*}

\subsubsection{Frequency Filtered Graph Analysis in the Vertex Domain}
In the spectral domain, analyzing the behavior of genetic signals reveals essential insights into spatial variability changes of cancer signals across weak, intermediate, and strong network connectivity. Notably, a consistent frequency magnitude between normal and cancer signals doesn't necessarily imply a static contribution from each gene within that frequency. Thus, examining the variability source (genes) in genetic signals for each biological condition in the vertex domain is necessary. Given the definition in (\ref{filter}), we focus on analyzing the low, medium, and high-pass filtered expression levels, denoted as  \({x}_{L}, {x}_{M}, {x}_{H}\). 

\textbf{Gene Significance: Graph Frequency-Based and in Isolation:} Firstly, we leverage the well-established WGCNA's Gene Significance measure \cite{horvath2008geometric} to define the significance of a gene in isolation. This measure is denoted as \(GS_j = \text{cor}(\mathbf{x_j}, \mathbf{T_i})\). As mentioned earlier, this metric determines the association between a variable and a specific biological trait, pinpointing genes whose expression levels exhibit significant differentiation with respect to the trait in question.

To identify genes with distinct spectral patterns, we modified the traditional approach. Instead of relying solely on original expression levels, we incorporated the genes' low, medium, and high frequency filtered expressions, \(\mathbf{\tilde{x}_{L}}, \mathbf{\tilde{x}_{M}}, \mathbf{\tilde{x}_{H}}\). This modification is expressed as $GS_j = \text{cor}(\mathbf{\tilde{x}_{kj}}, \mathbf{T_i})$, where \(k\) indicates the frequency range (\(L,M,H\)). 

The Gene Significance measure in isolation was obtained for each gene, along with three frequency-based gene significance levels for positive networks and three corresponding levels for negative networks. We averaged the three frequency-based measures within each network, resulting in two consolidated frequency-based significance levels for every gene.

\begin{figure}[htbp]
	\centering
	\includegraphics[width=2.5in]{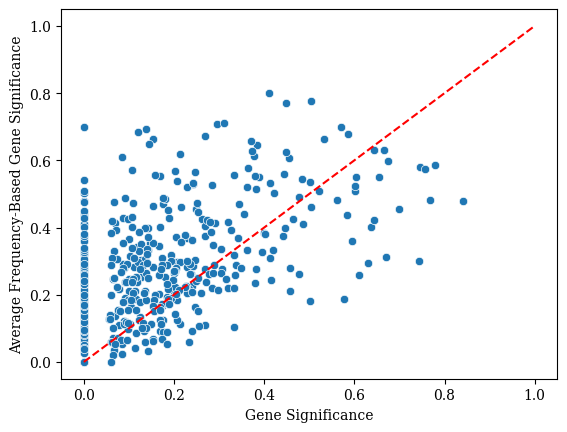}
	\caption{Association Between Gene Significance in Isolation and Graph Frequency-Based Significance.}
	\label{GeneSig}
\end{figure}

\begin{figure}[htbp]
	\centering
	\subfloat[\textrm{\footnotesize{Low-Frequency Filtered Signal}}]{\includegraphics[width=1.8in]{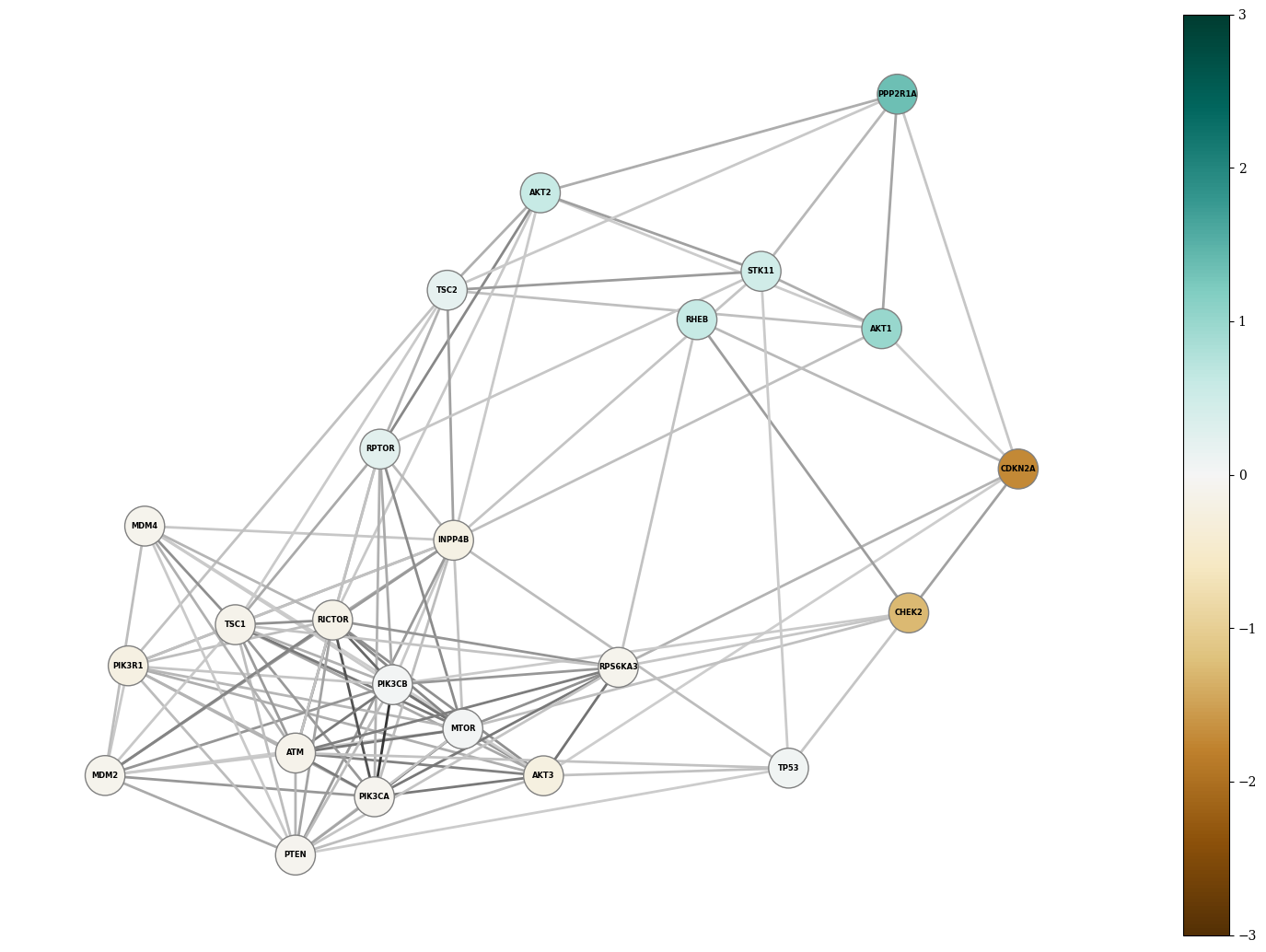}
		\label{Low}}
	\subfloat[\textrm{\footnotesize{Medium-Frequency Filtered Signal}}]{\includegraphics[width=1.8in]{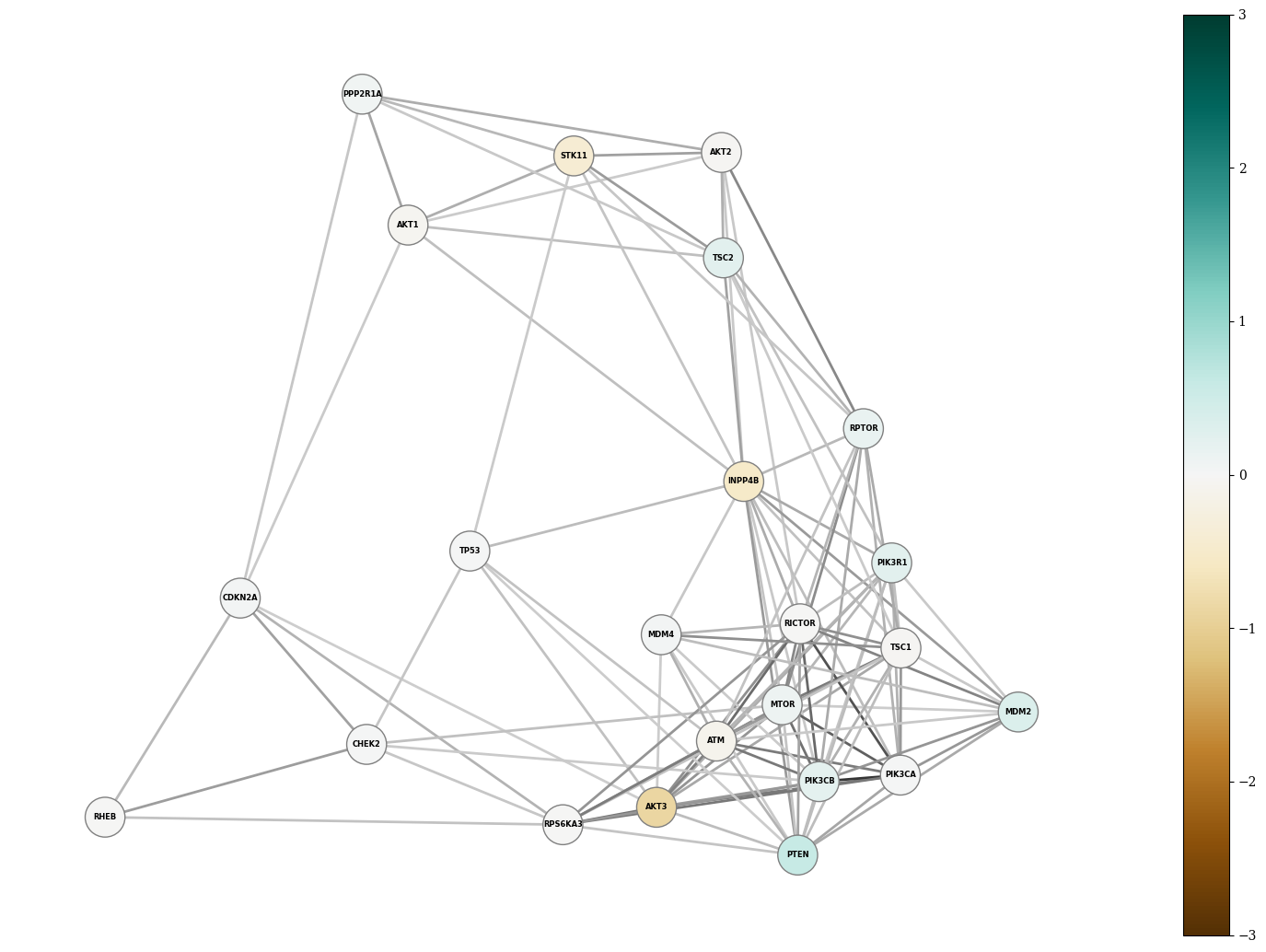}
	\label{Medium}}
	\hfil
	\subfloat[\textrm{\footnotesize{High-Frequency Filtered Signal}}]{\includegraphics[width=2in]{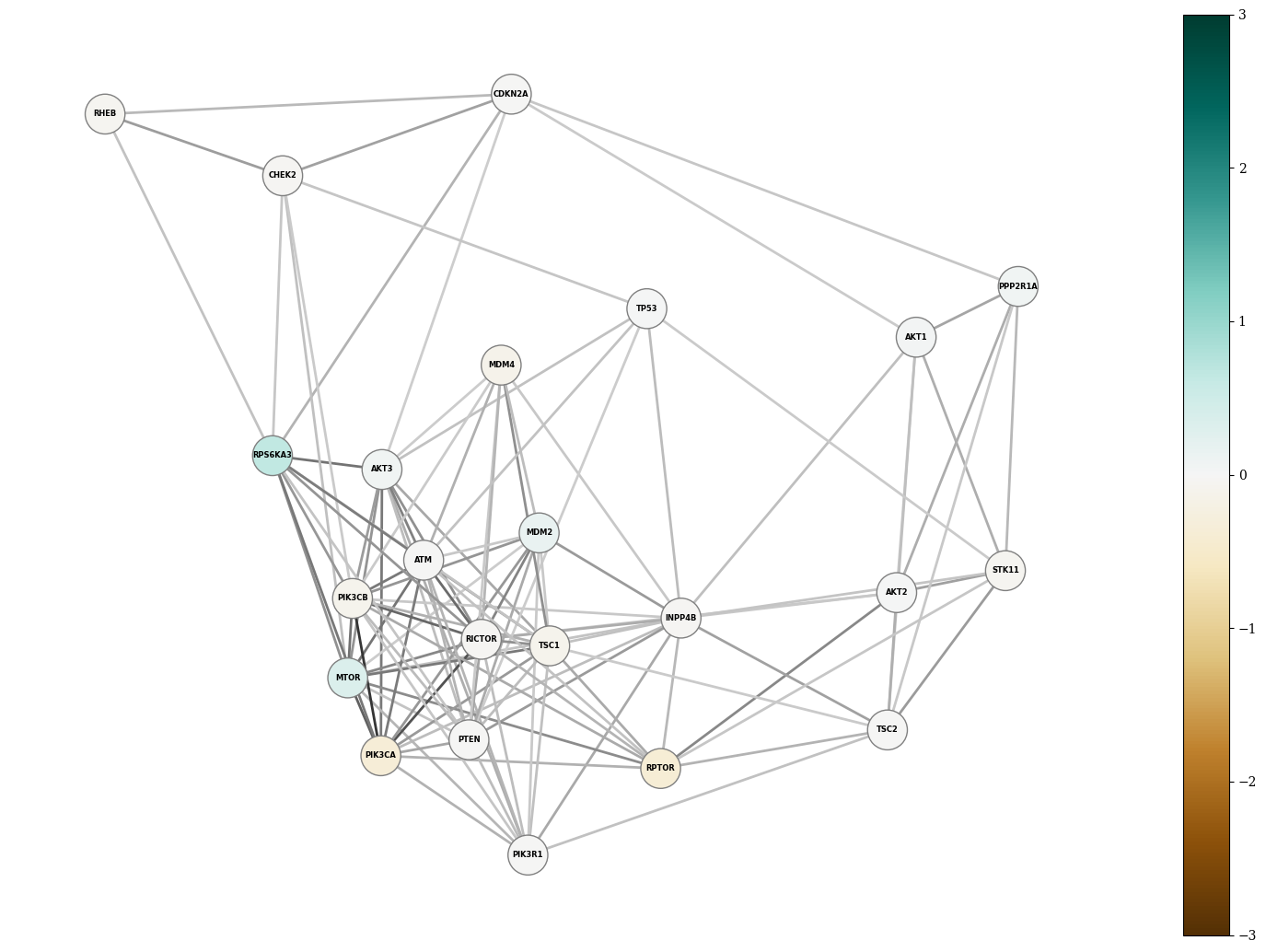}
		\label{High}}
        \hfil
	\caption{Average low (a), medium (b), and High-Frequency (c) Filtered Genetic Signal of Stomach Cancer Stage I Projected on The Positive Co-Expression Network.}
	\label{signal-decom}
\end{figure}

Subsequently, we measured the Pearson correlation coefficient between the gene significance's absolute values and the two separate frequency-based significance measures. \textbf{Our analysis revealed a statistically significant strong correlation (r=0.6, \(\mathbf{p < 0.05}\)) between the two significance measures}. This linear relationship is visually depicted in Figure \ref{GeneSig}.
\textbf{Gene Localization in Frequency Filtered-Expressions:} Upon examining the frequency-filtered expression levels in the vertex domain, distinct trends became evident concerning the connectivity of the genes in the networks. Specifically:
\begin{enumerate}
    \item Genes with a lower weighted degree, implying weak connectivity in the network, predominantly exhibited higher values in the low-frequency filtered signals. This pattern is evident in the average low-frequency filtered signal for stomach cancer stage I, as illustrated in Figure \ref{Low}.
    \item Genes with intermediate connectivity were the primary contributors to the medium-frequency filtered signal, as demonstrated in Figure \ref{Medium}.
    \item Hub genes, or those with the most connections within the network, manifested the highest values in high-frequency filtered signals.
\end{enumerate}
These observations align with the localization property of eigenvectors of the Laplacian, as detailed in prior research \cite{hata2017localization}.
The 40 co-expression networks were analyzed to confirm these findings, comprising 20 positive networks and 20 negative networks across five conditions (normal and cancer stages I through IV) for each of the studied cancer types. We calculated the average correlation between each gene's weighted degree and average absolute value in low, medium, and high-frequency filtered expressions. Our results showed:
\begin{enumerate}
    \item A strong average positive correlation (r=0.74, \(p<0.05\)) between the absolute values of the high-frequency filtered expression levels and the genes' weighted degrees defined in (\ref{wd}).
    \item A notable average negative correlation (r=-0.5, \(p<0.05\)) between the absolute values of the low-frequency filtered expression levels and the genes' connectivity or weighted degree.
    \item No significant correlation was observed between the genes' connectivity and the medium-frequency filtered expression levels.
\end{enumerate}
\textbf{Evaluation of Cancer Graph Frequency Analysis and Signature}:
To evaluate the ability of graph frequency analysis to extract informative features distinguishing normal and cancer stages gene expressions, we employed the F-statistics score, defined as the ratio of between-group to within-group variance. Table \ref{f-stat} shows the average scores of the genes when using first their original expression levels, second using their positive-networks based frequency filtered expressions $\mathbf{x^{\textbf{+}}_{k}}$ and negative -networks based frequency filtered expressions $\mathbf{x^{\textbf{-}}_{k}}$. Notably, all considered scores for this average held \(p\)-values below 0.05.
 The F-statistic measures differences between group means relative to variability within the groups. A high F-score indicates a feature strongly discriminates between groups. Table \ref{f-stat} unveils a salient observation: \textbf{gene features in all frequency-filtered genetic signals, both on positive and negative networks, possess markedly higher F-statistic scores than the original gene expression levels \(\mathbf{x_i}\)}. Intriguing characteristics of the genes emerge when we discern between normal and different cancer stages (I-IV) based on these frequency ranges. 
 
 Frequency analysis in the spectral domain revealed key changes in cancer signals' spatial variability. However, here, we can also see the importance of frequency analysis in the vertex domain. For example, lung cancer did not have a significant change in high-frequency amplitude between normal and cancer stages yet showed a very high average F-score of 378.7 for high-frequency filtered signals (Table \ref{f-stat}). Indicating substantial changes in genes contributing to the spatial variability between neighboring nodes (high-frequency amplitude). 

\begin{figure*}[htbp]
	\centering
	\subfloat[\textrm{\footnotesize
{Breast Gene Expression Levels}}]{\includegraphics[width=1.5in]{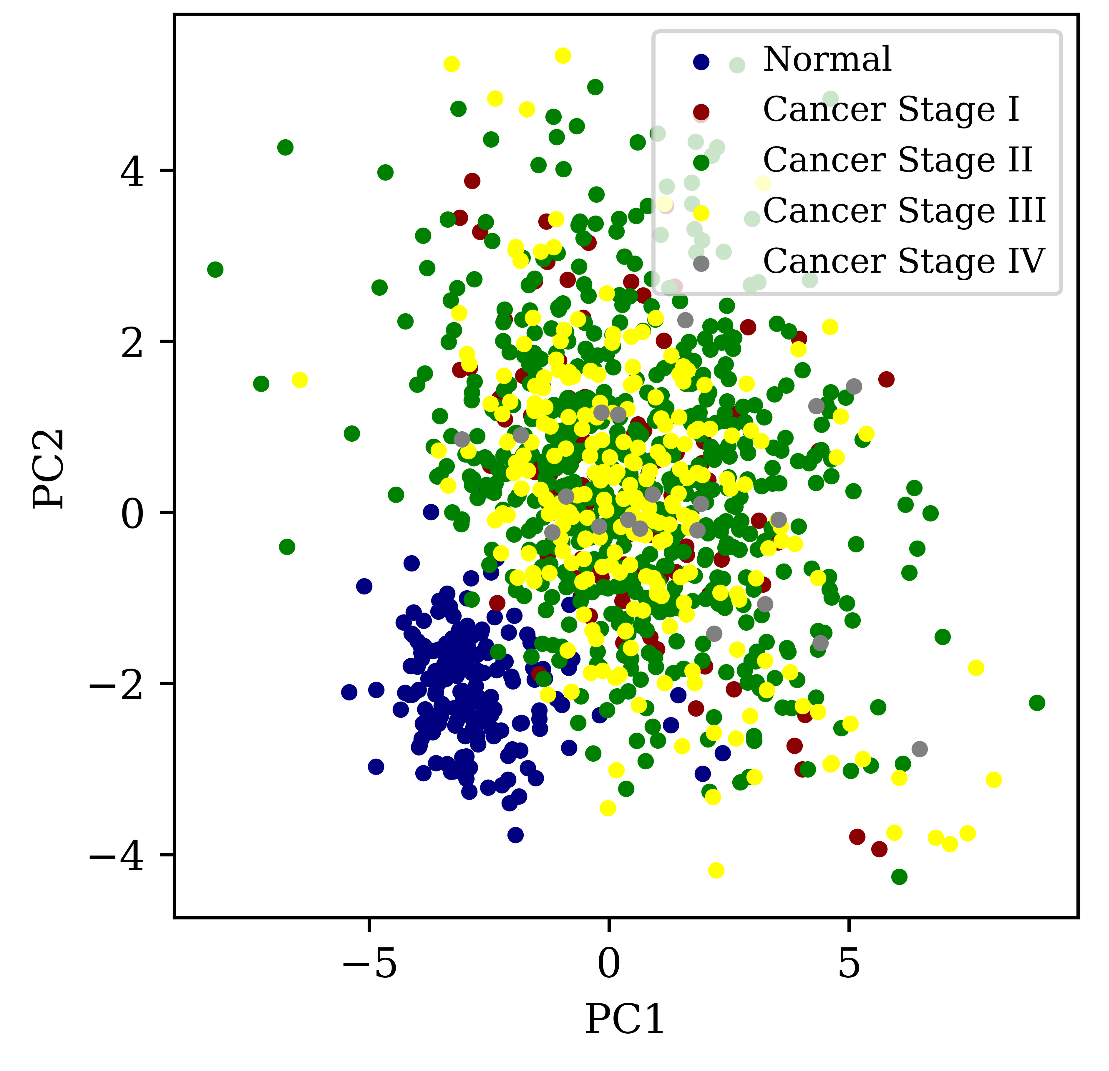}%
		\label{fig:A1}}
	\hfil
	\subfloat[\textrm{\footnotesize
{Breast High-Frequency Filtered Genetic Signals on Positive Networks}}]{\includegraphics[width=1.5in]{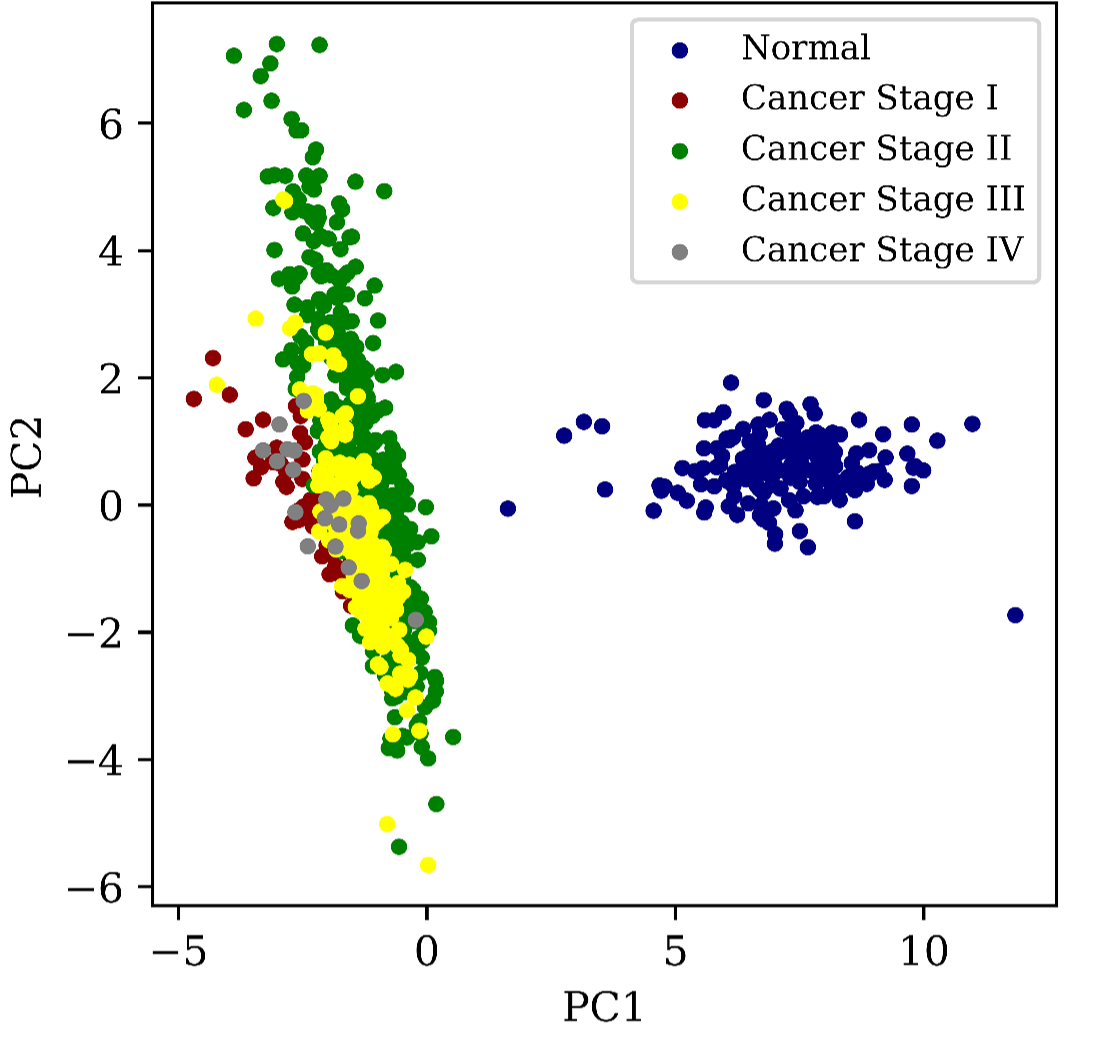}%
		\label{fig:A2}}
	\hfil
	\subfloat[\textrm{\footnotesize
{Breast Medium-Frequency Filtered Genetic Signals on The Negative Networks}}]{\includegraphics[width=1.5in]{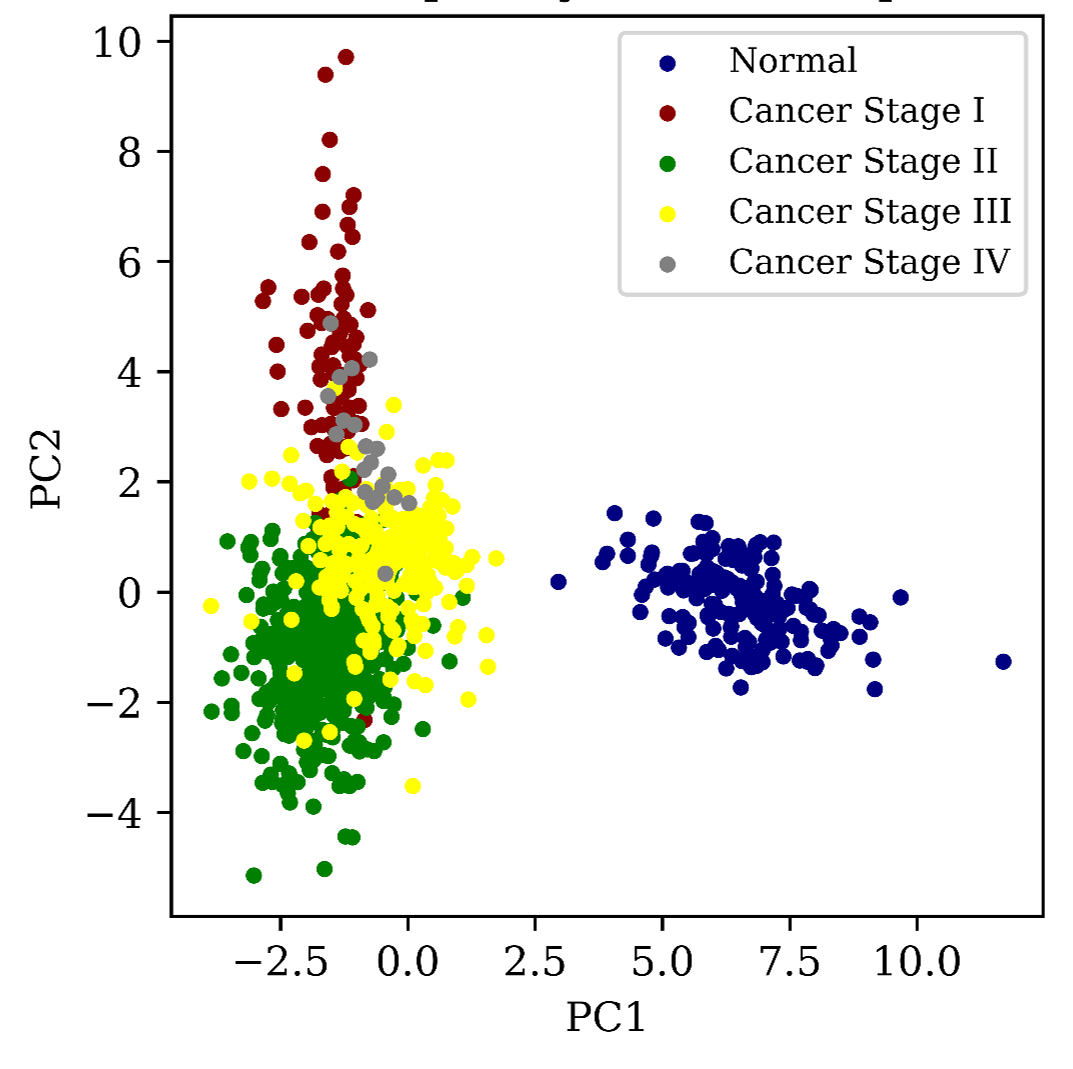}%
		\label{fig:A3}}

	\subfloat[\textrm{\footnotesize
{Colon Gene Expression Levels}}]{\includegraphics[width=1.5in]{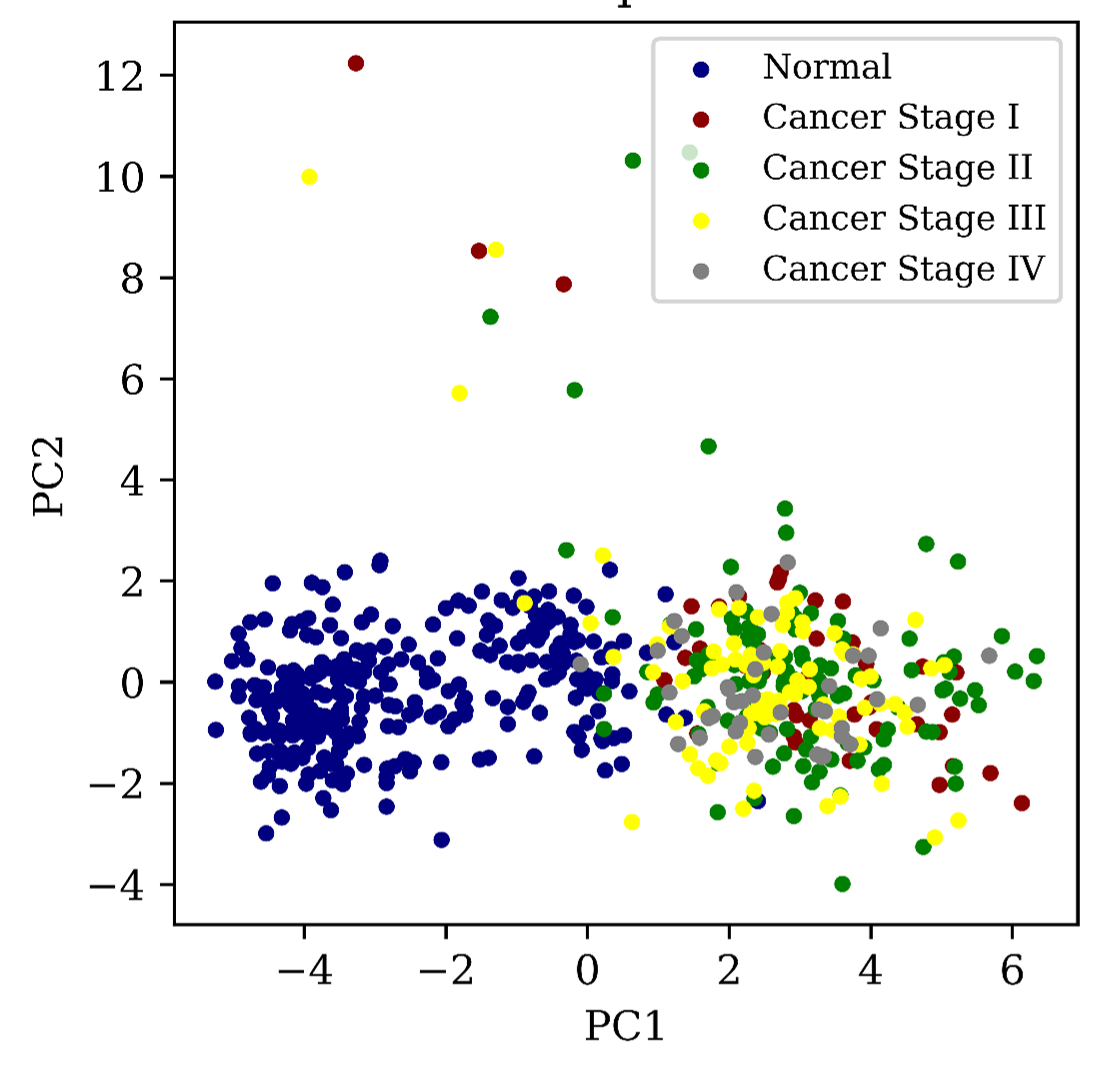}%
		\label{fig:B1}}
	\hfil
	\subfloat[\textrm{\footnotesize
{Colon Medium-Frequency Filtered Genetic Signals on Positive Networks}}]{\includegraphics[width=1.5in]{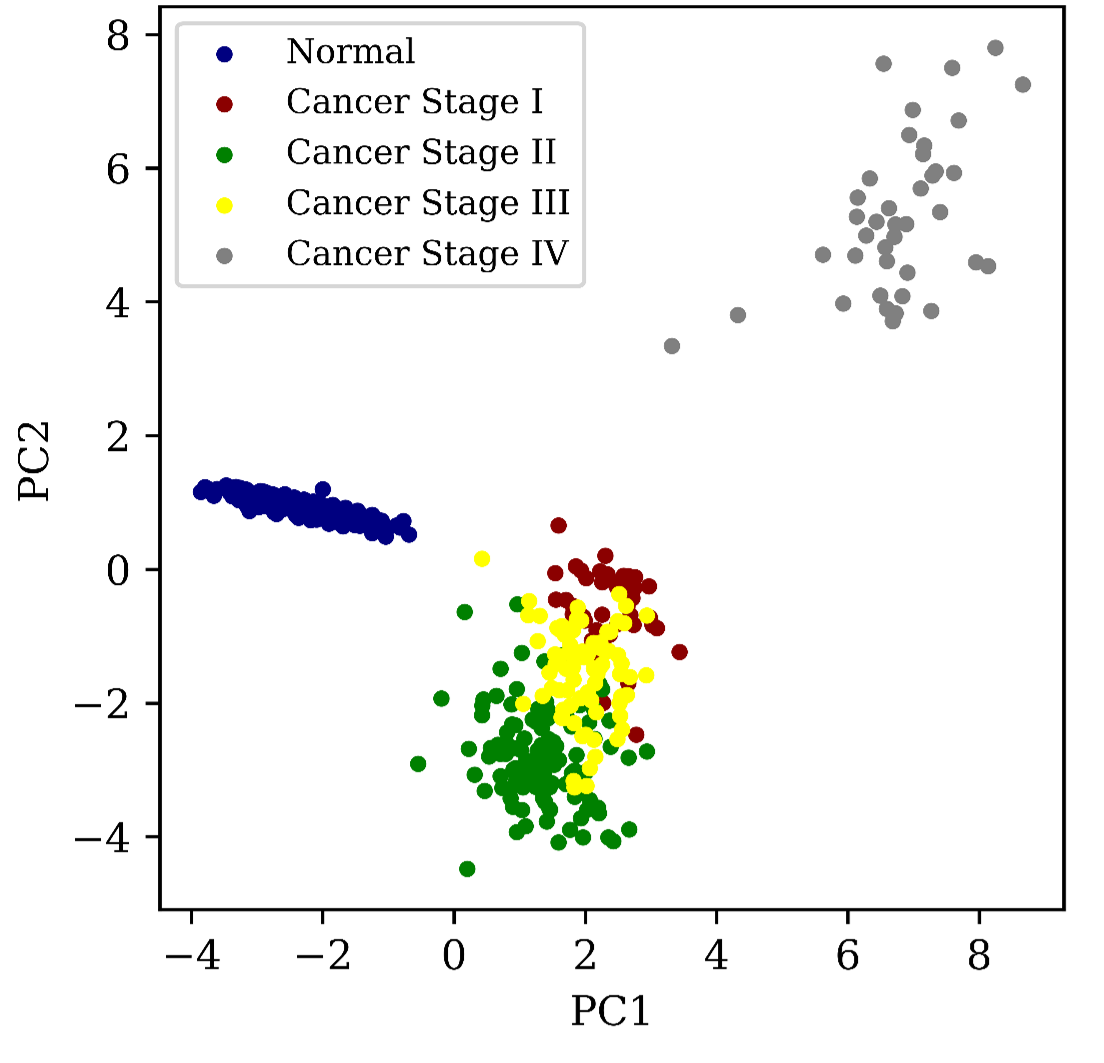}%
		\label{fig:B2}}
	\hfil
	\subfloat[\textrm{\footnotesize
{Colon Low-Frequency Filtered Genetic Signals on Negative Networks}}]{\includegraphics[width=1.5in]{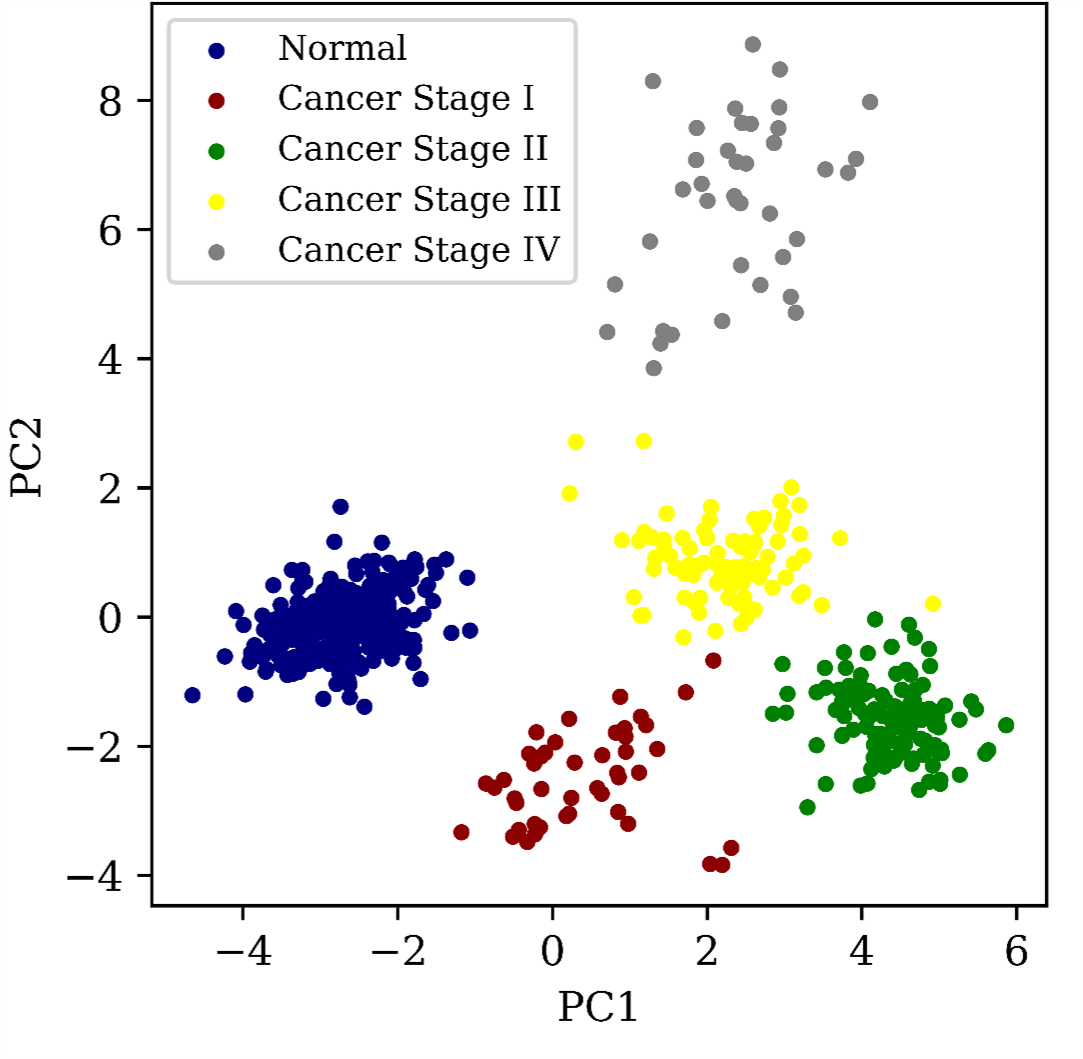}%
		\label{fig:B3}}

	\subfloat[\textrm{\footnotesize
{Lung Gene Expression Levels}}]{\includegraphics[width=1.5in]{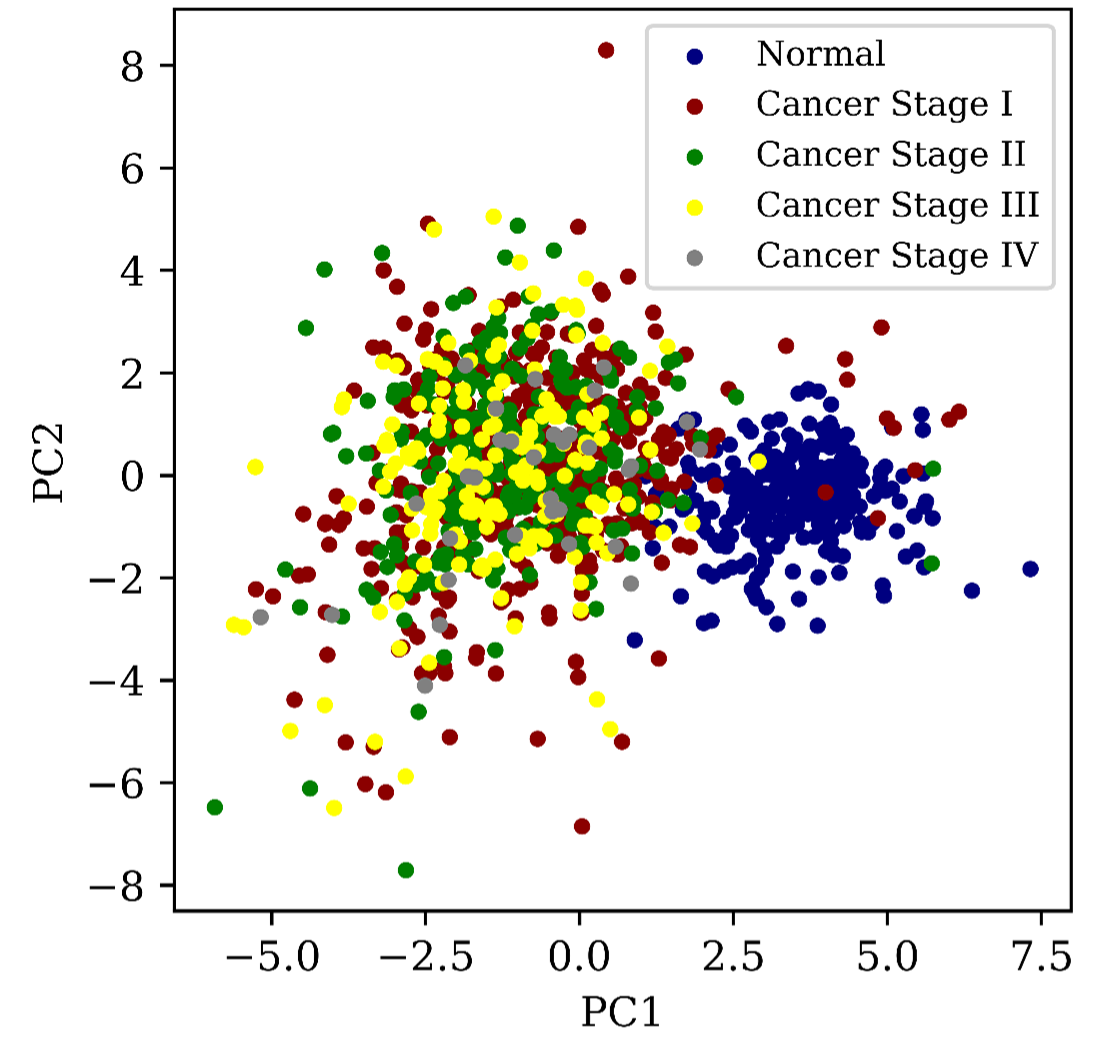}%
		\label{fig:C1}}
	\hfil
	\subfloat[\textrm{\footnotesize
{Lung Medium-Frequency Filtered Genetic Signals on Positive Networks}}]{\includegraphics[width=1.5in]{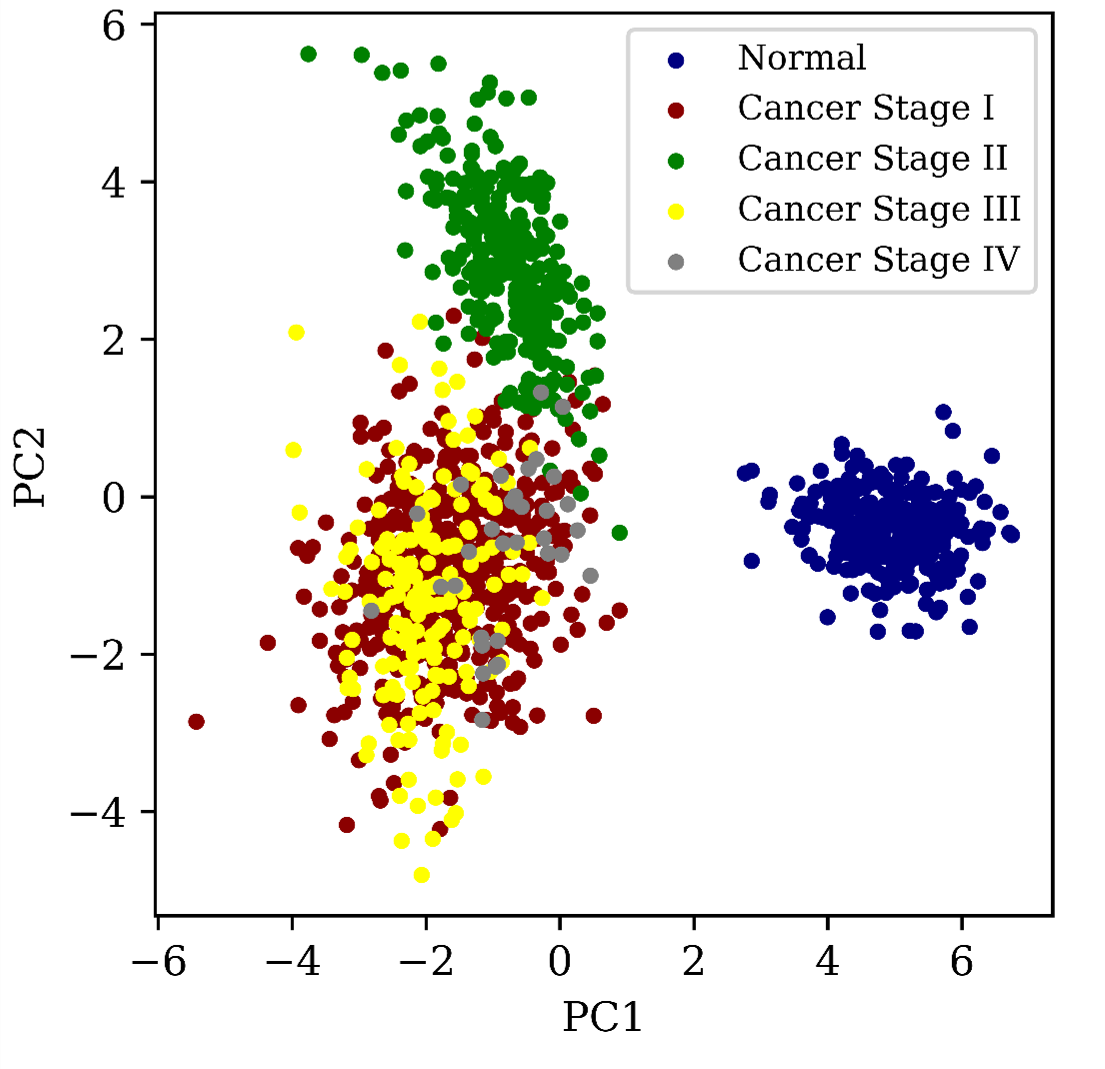}%
		\label{fig:C2}}
	\hfil
	\subfloat[\textrm{\footnotesize
{Lung Low-Frequency Filtered Genetic Signals on Negative Networks}}]{\includegraphics[width=1.5in]{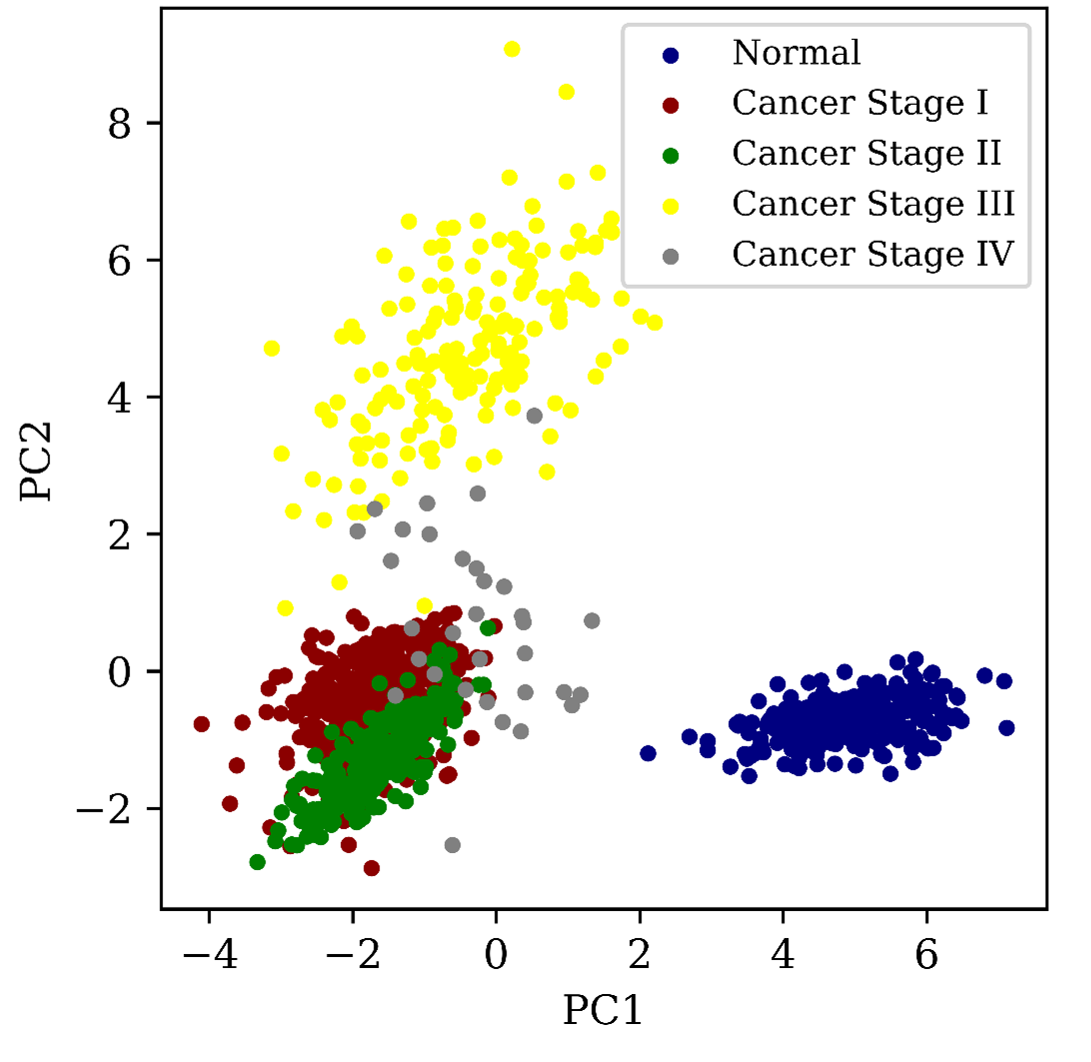}%
		\label{fig:C3}}

	\subfloat[\textrm{\footnotesize
{Stomach Gene Expression Levels}}]{\includegraphics[width=1.5in]{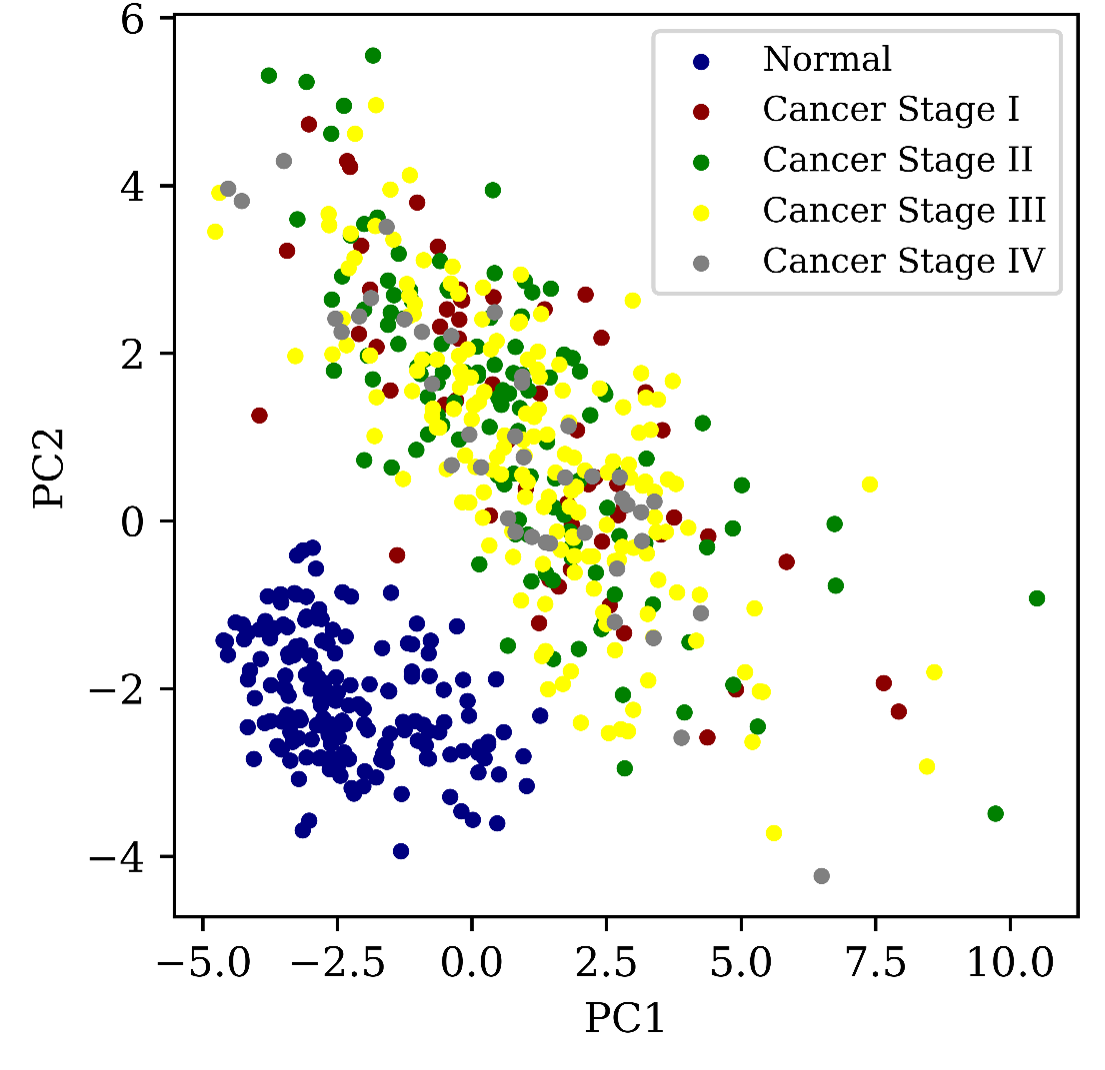}%
		\label{fig:D1}}
	\hfil
	\subfloat[\textrm{\footnotesize
{Stomach High-Frequency Filtered Genetic Signals on Positive Networks}}]{\includegraphics[width=1.5in]{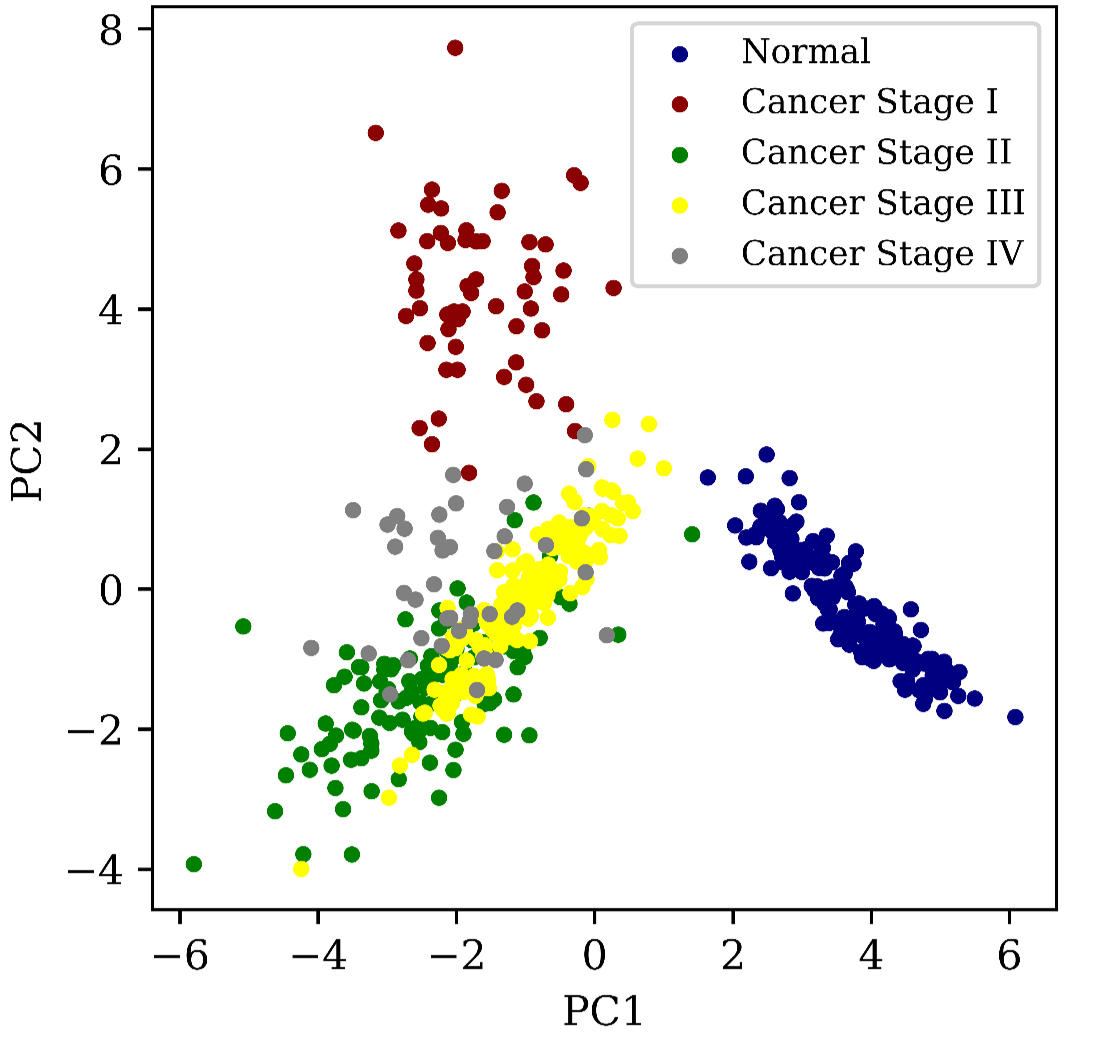}%
		\label{fig:D2}}
	\hfil
	\subfloat[\textrm{\footnotesize
{Stomach Low-Frequency Filtered Genetic Signals on Negative Networks}}]{\includegraphics[width=1.5in]{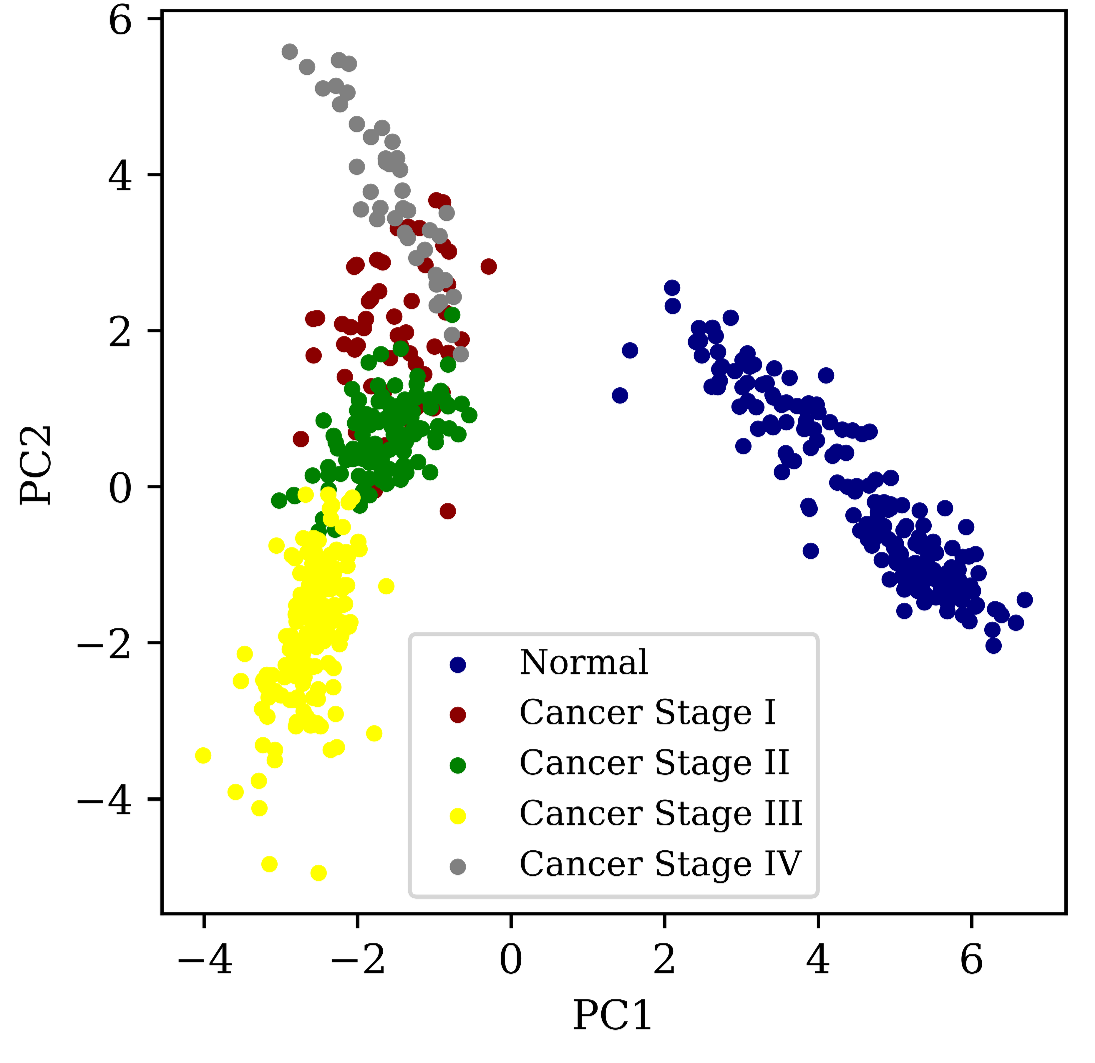}%
		\label{fig:D3}}
	
	\caption{Visualization of the First and Second Principal Components for Original and Frequency-Filtered Gene Expressions across Various Tissue Traits. The figures present scatter plots illustrate the first principal components (PC1 and PC2) extracted from the original and the frequency-filtered gene expression datasets, showcasing the highest F-statistic scores. The visualizations span four distinct tissue types: breast, colon, lung, and stomach, ordered from top to bottom rows.}
	\label{PCS}
\end{figure*}

Filtered signals across different frequency ranges exhibited varying strength levels for distinguishing different cancer types, as quantified by their F-scores as shown in Table \ref{f-stat}. Specifically, colon and lung cancers demonstrated higher F-scores for medium-frequency filtered signals on positive networks (600 and 820.5) and for low-frequency filtered signals on negative networks (718.6 and 855.8). In contrast, stomach cancer was better distinguished by high-frequency filtered signals on both positive and negative networks, with F-scores of 240 and 465, respectively. A notable F-score of 445.6 was also observed for stomach cancer using low-frequency filtered signals on negative networks. For breast cancer, high-frequency filtered signals on positive and negative networks showed a distinctive pattern with a comparatively high F-score of 1250 and 668, respectively. Therefore, different ranges of frequency-filtered gene expression levels enhance the discrimination ability of genes between normal and cancer stages. To visualize these observations, we extracted the first principal components from the datasets \textbf{X}, \textbf{$\mathbf{X^{\textbf{+}}_{k}}$}, \textbf{$\mathbf{X^{\textbf{-}}_{k}}$}, specifically from those frequency ranges that recorded the highest scores across both network types. The resulting visual representations are illustrated in Figure \ref{PCS}.

Another key observation is that \textbf{frequency-filtered gene signals on negative networks consistently yielded higher scores than their positive network counterparts}. This can be rationalized by the core principle of the graph Fourier transform and frequency analysis, which determine the spatial variability within the genetic signal across networks. By definition, such variability is anticipated to be more pronounced in networks defined by negative associations.
\begin{table}[htbp]
	\centering
	\caption{Average F-Statistic Score of The Genes Using Original and Frequency-Filtered  Expression Levels.\label{f-stat}}
	\begin{tabular}{|c||c|c|c|c|}
		\hline
		\textbf{Gene Values} & \textbf{Breast} & \textbf{Colon} & \textbf{Lung} & \textbf{Stomach} \\
		\hline
		\textbf{x} & 43.4 & 100 & 75.6 & 42.9 \\
		\hline
		\textbf{$\mathbf{x^{\textbf{+}}_{L}}$} & 288 & 308 & 313.8 & 142 \\
		\hline
		\textbf{$\mathbf{x^{\textbf{+}}_{M}}$} & 362 & \textbf{600} & \textbf{820.5} & 204 \\
		\hline
		\textbf{$\mathbf{x^{\textbf{+}}_{H}}$} & \textbf{1250} & 257 & 378.7 & \textbf{240} \\
		\hline
		\textbf{$\mathbf{x^{\textbf{-}}_{L}}$} & 414.7 & \textbf{718.6} & \textbf{855.8} & 445.6 \\
		\hline
		\textbf{$\mathbf{x^{\textbf{-}}_{M}}$} & \textbf{668.2} & 706.8 & 729 & 271 \\
		\hline
		\textbf{$\mathbf{x^{\textbf{-}}_{H}}$} & 248.9 & 303 & 350.7 & \textbf{465} \\
		\hline
	\end{tabular}
\end{table}

\textbf{Key Cancer Biomarker Genes in High-Frequency Filtered Expressions}: In our analysis of cancer genetic signals, key patterns emerged from genes across multiple cancer types. An important observation on the \textbf{mTOR} encoding gene (MTOR) is that it exhibited pronounced positive values in high-frequency filtered cancer genetic signals on positive co-expression networks across all analyzed cancer types. This indicates the over-expression of MTOR relative to its close neighboring genes in the co-expression network, a trend absent in normal high-frequency filtered genetic signals.
    
Moreover, the \textbf{RPS6KA3} gene, also known as Ribosomal Protein S6 Kinase A3, exhibited distinct over-expression in all stomach cancer stages, peaking at cancer stage I. This pattern was mirrored in colon cancer stage I but not in its advanced stages. Remarkably, its over-expression surpassed that of the mTOR gene in both stomach and colon cancers. When analyzing high-frequency filtered signals derived from negative network eigenvectors, the \textbf{PPP2R1A} gene and \textbf{AKT1} gene (integral to the PI3K/AKT/mTOR pathway) both displayed elevated positive values in lung, breast, and colon cancers. Conversely, the \textbf{INPP4B} gene presented pronounced negative values (indicative of under-expression) in lung cancer stages.

While multiple genes exhibited unique behaviors in cancer, our discussion emphasizes genes well-known for their biological significance and consistent patterns across multiple cancer types. It is essential to state that this study primarily demonstrates distinctive spectral characteristics of gene expressions. Further studies are required for a comprehensive biological interpretation of these patterns.
\section{Discussion}
\subsubsection{Progression of Cancer Gene Co-Expression Networks} 
In our study,  the selected genes were either oncogenes, which promote cell proliferation, or tumor suppressor genes, which regulate the cell cycle and critically control cell growth and division. These genes, instrumental in the P53 and PI3K pathways, and the shared complex co-regulation between the two pathways result in a balance in normal cells between cell cycle and growth, thereby maintaining cellular homeostasis. Notably, these genes are interconnected through functional interactions in our constructed networks, defined by statistically significant distance correlation between the genes, reflecting their linear and non-linear complex interactions.

In investigating gene co-expression networks in cancer, we found marked differences in connectivity patterns. Specifically, stomach and colon cancer-positive networks show increased functional connectivity as cancer initiates, in contrast to a decline in lung and breast cancer. All cancer types significantly had a reduced negative network connectivity, especially colon and stomach cancer. Increased eigenvalue variance of colon and breast cancer positive co-expression networks highlighted an increase in the heterogeneity of the level of connectivity of the genes. 

This points to disruptions in key genes within the PI3K and P53 pathways, marked by suppressed inhibitory regulation and dynamic shifts in co-activation interactions, especially in stomach and colon cancers. Our observations resonate with existing literature on the dysregulation of the P53 and PI3K pathways in cancer \cite{liu2012identifying, rothe2014plasma, singh2002p53}. Importantly, here we highlight the notable loss of negative interactions among studied genes. This highlights the diminished regulation of the PI3K pathway, resulting in uncontrolled cellular proliferation.
\subsubsection{Cancer Signals in The Spectral Domain} 
Our results demonstrate a compelling association between the weighted degree and eigenvalue distribution, underscoring the value of the Laplacian eigenvectors and eigenvalues to capture characteristics of the co-expression network topology. Therefore, this has motivated us further to analyze cancer genetic signals in the spectral domain. 

We observed distinctive global and cancer-specific characteristics in the spectral domain of cancer genetic signals. All types of cancer showed more similar expression levels between genes with weak connectivity in the positive co-expression networks, identified by the decrease in the low-frequency magnitude of cancer signals. Genes with weak connectivity might be less influenced by the network's underlying activating regulation dynamics. External factors like the tumor microenvironment might influence their expression in cancer cells \cite{HANAHAN2011646} rather than intricate intracellular signaling. This environment's metabolic and signaling molecules may encourage a similar set of gene expression patterns. 

Although overall gene connectivity in positive co-expression networks decreased in breast and stage I lung cancer, genes with intermediate and high connectivity showed more similar expression levels, as evidenced by reduced medium and high-frequency components. Additionally, stomach cancer presented another intriguing trend. A pronounced rise in the DC amplitude within the positive networks, coupled with diminished amplitudes in the high and low-frequency signal components. These findings are accompanied by increased network connectivity. Therefore, they suggest a strong activating regulatory mechanism among the selected twenty-three genes with similar high expression levels in stomach cancer. Such organized behavior in stomach, breast, and lung cancer stage I suggests the dysregulation of other regulatory mechanisms the genes used to participate in. Indicating the imbalance in cell functions; while normal cells maintain diversity in expression among highly connected genes for functional robustness and specialization, cancer disrupts this harmony, driving convergence in high expression patterns for survival, proliferation, and other oncogenic processes.

Colon cancer, however, displayed a different dynamic. We noted a decline in the overall DC amplitude of positive networks of cancer stage I, accompanied by an increase in the high-frequency genetic signal component. Given the observed high variance in the eigenvalues of the positive co-expression network—with most of these values falling below the median—we infer the emergence of hub genes during colon cancer initiation. These genes likely manifest higher expression levels than strongly connected genes, leading to pronounced spatial variability among neighboring genes, as captured by the high-frequency amplitude. However, as the disease progresses, this disparity weakens, characterized by reduced connectivity and increased high-frequency amplitudes in cancer stage II. Such trends indicate the chosen genes' inconsistent regulation mechanism in the evolution of colon cancer.

The spectral analysis of cancer genetic signals on negative networks also highlights the limitations of relying solely on gene network topology to infer cancer cellular processes. For instance, despite a marked reduction in connectivity within the lung cancer negative co-expression network, we observed a significant upsurge in the magnitude of high-frequency components during early cancer stages. This suggests an elevated dissimilarity in expression levels among closely connected genes, leading to rapid fluctuations in genetic signals across the network.

A consistent trend across all examined cancer types was the diminished magnitude of low-frequency components in genetic signals over the negative networks compared to their normal counterparts. This reduction can be attributed to various factors. One plausible explanation is the homogenizing pressures imposed by the tumor micro-environment, compelling specific genes to converge in their expression profiles and resist cell death. 

Interestingly, the minimal spatial variability observed in the genetic signals of lung and breast cancer within intermediate connections could indicate a more dominant influence of cancer's metabolic landscape or the tumor micro-environment than stomach and colon cancer. The fact that the network connectivity in breast and lung cancers did not decline to the extent observed in colon and stomach cancers further reinforces this conclusion.

These behaviors of cancer signals in the spectral domain, can be crucial in offering potential explanations for drug resistance in cancer and providing avenues to model and regulate uncontrolled cellular proliferation. Nevertheless,  differentiating the magnitudes of spectral components of genetic signals across networks with diverse topological structures can be restrictive. Such an approach primarily offers global insights into the gene expression patterns relative to different scales of connectivity within the network. It might overlook the possibility that while amplitudes remain similar, the sources of spatial variability within the network could change. 
\subsubsection{Frequency Filtered Graph Analysis in the Vertex Domain}
Our findings reveal a strong association between a gene's individual significance, derived from its differentiated expression levels, and the differentiation of its spectral contribution patterns. This correlation underscores the potency of incorporating gene expression data into spectral analysis. Such integration can discern spectral patterns pinpointing pivotal genes within isolation and in cancer systems, emphasizing the comprehensive nature of graph frequency analysis.

Furthermore, as revealed by our analysis, the average frequency-filtered expression levels demonstrate the graph frequency analysis's capability to spotlight both peripherally connected genes and core network components. This observation is rooted in the heterogeneous weighted degree of the genes resulting in the localization property of the Laplacian eigenvectors \cite{hata2017localization}. However, it is essential to acknowledge that the values in the frequency-filtered signals are influenced not solely by a gene's connectivity strength but also by its relative expression level within the network. Thus, visualization of network topology is necessary for accurately discerning less connected genes. 

Nonetheless, this approach offers valuable insights into a gene's position within the network, aiding in identifying either over-expressed or under-expressed network core genes, therefore central to the underlying biological process. Several genes crucial to cancer biology in our study showed distinct expression patterns within their respective networks. The MTOR gene, a central player in cell growth and proliferation, was consistently over-expressed across all cancer types. This supports the well-established role of MTOR in various human cancers \cite{forbes2010cosmic}, emphasizing its heightened activity during oncogenic processes.

In stomach cancer, the gene RPS6KA3 was notably over-expressed compared to its network neighbors. The Human Protein Atlas data reinforces this observation, identifying RPS6KA3's encoded protein as a significant marker in stomach cancer pathology.
The AKT1 gene, crucial for inhibiting apoptosis, showed pronounced over-expression in lung and colon cancers. This aligns with its known mutations in these cancer types \cite{sanchez2018oncogenic} and its pivotal role in negating the effects of tumor suppressor genes. Our analysis also highlighted the PPP2R1A gene, encoding the regulatory subunit of PP2A. While traditionally a tumor suppressor, its enhanced expression and central network position in lung, breast, and colon cancers suggest a possible oncogenic shift in these contexts \cite{bian2014synthetic}.

Lastly, INPP4B, an essential PI3K pathway tumor suppressor, was significantly under-expressed in lung cancer. Its central position in negative co-expression networks and relative under-expression suggests a complex interplay with surrounding genes, potentially upregulating its neighbors \cite{stjernstrom2014alterations}.
Our findings offer a nuanced understanding of gene expressions within cancer networks, emphasizing the significance of over-expressed and under-expressed genes in the core of these networks. These results, combined with the elevated average f-statistic scores of genes using spectral features, underscore the efficacy of graph frequency analysis in discerning topological and signal gene features within cancer processes.
\subsubsection{Implications and Limitations of Findings}
Our findings suggest that graph frequency analysis is critical in analyzing cancer systems and extracting key genes within the cellular systems. We found marked evidence of cancer characteristics patterns in the spectral domain. Therefore, graph frequency analysis can further aid therapeutic drug design by assessing drug impacts on cancer cells and understanding subsequent spectral components. Additionally, it opens avenues for modeling cancer co-expression network dynamics and pinpointing gene targets for cancer drug studies. Finally, differential spectral analysis presents a promising application for differential co-expression analysis in cancer because it effectively integrates genetic signals and network topology. 

Challenges in this approach persist in understanding cancer's spectral behavior biologically. Machine learning methods such as clustering can be further used for improved identification of key cancer genes based on their features in the graph frequency filtered signals. Moreover, while we employed Laplacian eigendecomposition in designing the graph frequency filters, it is computationally intensive for large networks.
\section{Conclusions}
 We used graph frequency analysis methods to analyze gene expression profiles and co-expression networks in normal and cancer stages (I-IV) for four types of cancer. We validated that the frequency content of gene expression profiles can significantly distinguish between normal and cancer systems. Additionally, it can successfully identify key cancer genes by incorporating their expression levels and location in the network. Our findings also showed evidence of cancer inter-tumor heterogeneity in the spectral domain. Different ranges of frequencies were better at discriminating different types of cancer samples. We conclude that the graph spectral properties of cancer signals defined on gene co-expression networks can be highly beneficial in differential co-expression analysis. Finally, in our future research, motivated by the results and models in \cite{SalasGonzalezKuruogluRuiz+2009}\cite{JMLR:v17:14-150}\cite{refId0}\cite{SAFFARIMIANDOAB2021115197}, we will incorporate also the statistical distribution information about the data in the graph model.

%
\section*{Acknowledgments}
This work is supported by Shenzhen Science and Technology Innovation Commission under Grant JCYJ202205301430\\02005 and Tsinghua Shenzhen International Graduate School Start-up fund under Grant QD2022024C. 

\bibliographystyle{IEEEtran}
\bibliography{refs}

\begin{thebibliography}{10}
\providecommand{\url}[1]{#1}
\csname url@samestyle\endcsname
\providecommand{\newblock}{\relax}
\providecommand{\bibinfo}[2]{#2}
\providecommand{\BIBentrySTDinterwordspacing}{\spaceskip=0pt\relax}
\providecommand{\BIBentryALTinterwordstretchfactor}{4}
\providecommand{\BIBentryALTinterwordspacing}{\spaceskip=\fontdimen2\font plus
\BIBentryALTinterwordstretchfactor\fontdimen3\font minus
  \fontdimen4\font\relax}
\providecommand{\BIBforeignlanguage}[2]{{%
\expandafter\ifx\csname l@#1\endcsname\relax
\typeout{** WARNING: IEEEtran.bst: No hyphenation pattern has been}%
\typeout{** loaded for the language `#1'. Using the pattern for}%
\typeout{** the default language instead.}%
\else
\language=\csname l@#1\endcsname
\fi
#2}}
\providecommand{\BIBdecl}{\relax}
\BIBdecl

\bibitem{HANAHAN2011646}
D.~Hanahan and R.~A. Weinberg, ``Hallmarks of cancer: The next generation,''
  \emph{Cell}, vol. 144, pp. 646--674, 2011.

\bibitem{holohan2013cancer}
C.~Holohan, S.~Van~Schaeybroeck, D.~B. Longley, and P.~G. Johnston, ``Cancer
  drug resistance: an evolving paradigm,'' \emph{Nature Reviews Cancer},
  vol.~13, no.~10, pp. 714--726, 2013.

\bibitem{Sun2015}
X.~{X}iao Sun and Q.~Yu, ``Intra-tumor heterogeneity of cancer cells and its
  implications for cancer treatment,'' \emph{Acta Pharmacologica Sinica},
  vol.~36, pp. 1219--1227, 2015.

\bibitem{Wang2009}
Z.~Wang, M.~Gerstein, and M.~Snyder, ``{RNA}-seq: a revolutionary tool for
  transcriptomics,'' \emph{Nature Reviews Genetics}, vol.~10, pp. 57--63, 2009.

\bibitem{Soneson2013}
C.~Soneson and M.~Delorenzi, ``A comparison of methods for differential
  expression analysis of rna-seq data,'' \emph{BMC Bioinformatics}, vol.~14,
  p.~91, 2013.

\bibitem{LONG2023}
F.~Long, H.~Ma, Y.~Hao, L.~Tian, Y.~Li, B.~Li, J.~Chen, Y.~Tang, J.~Li, L.~Deng
  \emph{et~al.}, ``A novel exosome-derived prognostic signature and risk
  stratification for breast cancer based on multi-omics and systematic
  biological heterogeneity,'' \emph{Computational and {Structural}
  {Biotechnology} {Journal}}, vol.~21, pp. 3010--3023, 2023.

\bibitem{tyson2001network}
J.~J. Tyson, K.~Chen, and B.~Novak, ``Network dynamics and cell physiology,''
  \emph{{Nature} {Reviews} {Molecular} {Cell} {Biology}}, vol.~2, no.~12, pp.
  908--916, 2001.

\bibitem{zhang2005general}
B.~Zhang and S.~Horvath, ``A general framework for weighted gene co-expression
  network analysis,'' \emph{Statistical {Applications} in {Genetics} and
  {Molecular Biology}}, vol.~4, no.~1, 2005.

\bibitem{horvath2008geometric}
S.~Horvath and J.~Dong, ``Geometric interpretation of gene coexpression network
  analysis,'' \emph{PLoS {Computational} {Biology}}, vol.~4, no.~8, p.
  e1000117, 2008.

\bibitem{sandhu2015graph}
R.~Sandhu, T.~Georgiou, E.~Reznik, L.~Zhu, I.~Kolesov, Y.~Senbabaoglu, and
  A.~Tannenbaum, ``Graph curvature for differentiating cancer networks,''
  \emph{Scientific Reports}, vol.~5, no.~1, p. 12323, 2015.

\bibitem{tesson2010diffcoex}
B.~M. Tesson, R.~Breitling, and R.~C. Jansen, ``Diffcoex: a simple and
  sensitive method to find differentially coexpressed gene modules,'' \emph{BMC
  Bioinformatics}, vol.~11, no.~1, pp. 1--9, 2010.

\bibitem{pettersen2022csdr}
J.~P. Pettersen and E.~Almaas, ``csd{R}, an {R} package for differential
  co-expression analysis,'' \emph{BMC {Bioinformatics}}, vol.~23, no.~1, pp.
  1--7, 2022.

\bibitem{west2012differential}
J.~West, G.~Bianconi, S.~Severini, and A.~E. Teschendorff, ``Differential
  network entropy reveals cancer system hallmarks,'' \emph{Scientific Reports},
  vol.~2, no.~1, p. 802, 2012.

\bibitem{6494675}
D.~I. Shuman, S.~K. Narang, P.~Frossard, A.~Ortega, and P.~Vandergheynst, ``The
  emerging field of signal processing on graphs: Extending high-dimensional
  data analysis to networks and other irregular domains,'' \emph{IEEE Signal
  Processing Magazine}, vol.~30, no.~3, pp. 83--98, 2013.

\bibitem{RICAUD2019474}
B.~Ricaud, P.~Borgnat, N.~Tremblay, P.~Gonçalves, and P.~Vandergheynst,
  ``Fourier could be a data scientist: From graph fourier transform to signal
  processing on graphs,'' \emph{Comptes Rendus Physique}, vol.~20, no.~5, pp.
  474--488, 2019, {Fourier} and the science of today / Fourier et la science
  d’aujourd’hui.

\bibitem{8347162}
A.~Ortega, P.~Frossard, J.~Kova{\v{c}}evi{\'{c}}, J.~M.~F. Moura, and
  P.~Vandergheynst, ``{Graph Signal Processing: Overview, Challenges, and
  Applications},'' \emph{Proceedings of the IEEE}, vol. 106, no.~5, pp.
  808--828, 2018.

\bibitem{DBLP:journals/corr/SegarraHR14}
S.~Segarra, W.~Huang, and A.~Ribeiro, ``Diffusion and superposition distances
  for signals supported on networks,'' \emph{CoRR}, vol. abs/1411.7443, 2014.

\bibitem{WANG2021100662}
W.~Wang, F.~Zhou, D.~B. Tay, and J.~Jiang, ``Gene selection for cancer
  detection using graph signal processing,'' \emph{Informatics in Medicine
  Unlocked}, vol.~25, p. 100662, 2021.

\bibitem{Huang2016}
W.~Huang, L.~Goldsberry, N.~F. Wymbs, S.~T. Grafton, D.~S. Bassett, and
  A.~Ribeiro, ``{Graph Frequency Analysis of Brain Signals},'' \emph{IEEE
  {Journal} on {Selected} {Topics} in Signal {Processing}}, vol.~10, no.~7, pp.
  1189--1203, oct 2016.

\bibitem{sanchez2018oncogenic}
F.~Sanchez-Vega, M.~Mina, J.~Armenia, W.~K. Chatila, A.~Luna, K.~C. La,
  S.~Dimitriadoy, D.~L. Liu, H.~S. Kantheti, S.~Saghafinia \emph{et~al.},
  ``Oncogenic signaling pathways in the cancer genome atlas,'' \emph{Cell},
  vol. 173, no.~2, pp. 321--337, 2018.

\bibitem{singh2002p53}
B.~Singh, P.~G. Reddy, A.~Goberdhan, C.~Walsh, S.~Dao, I.~Ngai, T.~C. Chou,
  O.~Pornchai, A.~J. Levine, P.~H. Rao \emph{et~al.}, ``p53 regulates cell
  survival by inhibiting pik3ca in squamous cell carcinomas,'' \emph{Genes \&
  {Development}}, vol.~16, no.~8, pp. 984--993, 2002.

\bibitem{Goldman326470}
M.~Goldman, B.~Craft, M.~Hastie, K.~Repe{\v c}ka, F.~McDade, A.~Kamath,
  A.~Banerjee, Y.~Luo, D.~Rogers, A.~N. Brooks, J.~Zhu, and D.~Haussler, ``The
  {UCSC Xena} platform for public and private cancer genomics data
  visualization and interpretation,'' \emph{BioRxiv}, 2019.

\bibitem{weinstein2013cancer}
J.~N. Weinstein, E.~A. Collisson, G.~B. Mills, K.~R. Shaw, B.~A. Ozenberger,
  K.~Ellrott, I.~Shmulevich, C.~Sander, and J.~M. Stuart, ``The cancer genome
  atlas pan-cancer analysis project,'' \emph{{Nature Genetics}}, vol.~45,
  no.~10, pp. 1113--1120, 2013.

\bibitem{lonsdale2013genotype}
J.~Lonsdale, J.~Thomas, M.~Salvatore, R.~Phillips, E.~Lo, S.~Shad, R.~Hasz,
  G.~Walters, F.~Garcia, N.~Young \emph{et~al.}, ``The genotype-tissue
  expression {(GTEx)} project,'' \emph{Nature Genetics}, vol.~45, no.~6, pp.
  580--585, 2013.

\bibitem{langfelder2008wgcna}
P.~Langfelder and S.~Horvath, ``{WGCNA}: an r package for weighted correlation
  network analysis,'' \emph{BMC {Bioinformatics}}, vol.~9, no.~1, pp. 1--13,
  2008.

\bibitem{szekely2009brownian}
G.~J. Sz{\'e}kely and M.~L. Rizzo, ``Brownian distance covariance,'' \emph{{The
  Annals of Applied Statistics}}, pp. 1236--1265, 2009.

\bibitem{hou2022distance}
J.~Hou, X.~Ye, W.~Feng, Q.~Zhang, Y.~Han, Y.~Liu, Y.~Li, and Y.~Wei, ``Distance
  correlation application to gene co-expression network analysis,'' \emph{BMC
  Bioinformatics}, vol.~23, no.~1, pp. 1--24, 2022.

\bibitem{karaaslanli2022scsgl}
A.~Karaaslanli, S.~Saha, S.~Aviyente, and T.~Maiti, ``{scSGL}: kernelized
  signed graph learning for single-cell gene regulatory network inference,''
  \emph{Bioinformatics}, vol.~38, no.~11, pp. 3011--3019, 2022.

\bibitem{chung1997spectral}
F.~R. Chung, \emph{Spectral Graph Theory}.\hskip 1em plus 0.5em minus
  0.4em\relax American Mathematical Soc., 1997, vol.~92.

\bibitem{ricaud2019fourier}
B.~Ricaud, P.~Borgnat, N.~Tremblay, P.~Gon{\c{c}}alves, and P.~Vandergheynst,
  ``Fourier could be a data scientist: From graph fourier transform to signal
  processing on graphs,'' \emph{Comptes Rendus Physique}, vol.~20, no.~5, pp.
  474--488, 2019.

\bibitem{fiedler1973algebraic}
M.~Fiedler, ``Algebraic connectivity of graphs,'' \emph{Czechoslovak
  Mathematical Journal}, vol.~23, no.~2, pp. 298--305, 1973.

\bibitem{tremblay2014graph}
N.~Tremblay and P.~Borgnat, ``Graph wavelets for multiscale community mining,''
  \emph{IEEE Transactions on Signal Processing}, vol.~62, no.~20, pp.
  5227--5239, 2014.

\bibitem{hata2017localization}
S.~Hata and H.~Nakao, ``Localization of {Laplacian} eigenvectors on random
  networks,'' \emph{Scientific Reports}, vol.~7, no.~1, p. 1121, 2017.

\bibitem{huang2016graph}
W.~Huang, L.~Goldsberry, N.~F. Wymbs, S.~T. Grafton, D.~S. Bassett, and
  A.~Ribeiro, ``Graph frequency analysis of brain signals,'' \emph{IEEE
  {Journal} of {Selected} {Topics} in {Signal} {Processing}}, vol.~10, no.~7,
  pp. 1189--1203, 2016.

\bibitem{liu2012identifying}
K.-Q. Liu, Z.-P. Liu, J.-K. Hao, L.~Chen, and X.-M. Zhao, ``Identifying
  dysregulated pathways in cancers from pathway interaction networks,''
  \emph{BMC Bioinformatics}, vol.~13, no.~1, pp. 1--11, 2012.

\bibitem{rothe2014plasma}
F.~Roth{\'e}, J.-F. Laes, D.~Lambrechts, D.~Smeets, D.~Vincent, M.~Maetens,
  D.~Fumagalli, S.~Michiels, S.~Drisis, C.~Moerman \emph{et~al.}, ``Plasma
  circulating tumor dna as an alternative to metastatic biopsies for mutational
  analysis in breast cancer,'' \emph{Annals of Oncology}, vol.~25, no.~10, pp.
  1959--1965, 2014.

\bibitem{forbes2010cosmic}
S.~A. Forbes, N.~Bindal, S.~Bamford, C.~Cole, C.~Y. Kok, D.~Beare, M.~Jia,
  R.~Shepherd, K.~Leung, A.~Menzies \emph{et~al.}, ``Cosmic: mining complete
  cancer genomes in the catalogue of somatic mutations in cancer,''
  \emph{Nucleic {Acids} {Research}}, vol.~39, no. suppl\_1, pp. D945--D950,
  2010.

\bibitem{bian2014synthetic}
Y.~Bian, R.~Kitagawa, P.~K. Bansal, Y.~Fujii, A.~Stepanov, and K.~Kitagawa,
  ``Synthetic genetic array screen identifies pp2a as a therapeutic target in
  mad2-overexpressing tumors,'' \emph{Proceedings of the National Academy of
  Sciences}, vol. 111, no.~4, pp. 1628--1633, 2014.

\bibitem{stjernstrom2014alterations}
A.~Stjernstr{\"o}m, C.~Karlsson, O.~J. Fernandez, P.~S{\"o}derkvist, M.~G.
  Karlsson, and L.~K. Thunell, ``Alterations of inpp4b, pik3ca and pakt of the
  pi3k pathway are associated with squamous cell carcinoma of the lung,''
  \emph{Cancer Medicine}, vol.~3, no.~2, pp. 337--348, 2014.

\bibitem{SalasGonzalezKuruogluRuiz+2009}
\BIBentryALTinterwordspacing
D.~Salas-Gonzalez, E.~E. Kuruoglu, and D.~P. Ruiz, ``Modelling and assessing
  differential gene expression using the alpha stable distribution,'' \emph{The
  International Journal of Biostatistics}, vol.~5, no.~1, 2009.
\BIBentrySTDinterwordspacing

\bibitem{JMLR:v17:14-150}
\BIBentryALTinterwordspacing
N.~Misra and E.~E. Kuruoglu, ``Stable graphical models,'' \emph{Journal of
  Machine Learning Research}, vol.~17, no. 168, pp. 1--36, 2016.
\BIBentrySTDinterwordspacing

\bibitem{refId0}
\BIBentryALTinterwordspacing
D.~Herranz, E.~E. Kuruoğlu, and L.~Toffolatti,
``An $\alpha$-stable approach to the study of the p(d) distribution of unresolved point sources in cmb sky maps,'' \emph{A\&A}, vol. 424, no.~3, pp. 1081--1096, 2004.
\BIBentrySTDinterwordspacing

\bibitem{SAFFARIMIANDOAB2021115197}
\BIBentryALTinterwordspacing
F.~Saffarimiandoab, R.~Mattesini, W.~Fu, E.~E. Kuruoglu, and X.~Zhang,
  ``Insights on features' contribution to desalination dynamics and capacity of
  capacitive deionization through machine learning study,''
  \emph{Desalination}, vol. 515, p. 115197, 2021.
\BIBentrySTDinterwordspacing

\end{thebibliography}
\section{Biographies} 
\vspace{11pt}
\vspace{-33pt}
\begin{IEEEbiography}
{Radwa Adel}
received a B.S. degree in Communication Engineering from the School of Electronic and Information Engineering, Beijing Jiaotong University. She received a M.S. degree in Data Science and Information Technology from Tsinghua University, Tsinghua-Berkeley Shenzhen Institute of Tsinghua Shenzhen Graduate School. 
\end{IEEEbiography}
\vspace{11pt}
\vspace{-33pt}
\begin{IEEEbiography}
{Ercan E. Kuruoğlu} received MPhil and PhD degrees in information engineering from the University of Cambridge, United Kingdom, in 1995 and 1998, respectively. In 1998, he joined Xerox Research Center Europe, Cambridge. He was an ERCIM fellow in 2000 with INRIA-Sophia Antipolis, France. In January 2002, he joined ISTI-CNR, Pisa, Italy where he became a Chief Scientist in 2020. Currently, he is a Full Professor at Tsinghua-Berkeley Shenzhen Institute since March 2022. He served as an Associate Editor for the IEEE Transactions on Signal Processing and IEEE Transactions on Image Processing. He was the Editor in Chief of Digital Signal Processing: A Review Journal between 2011-2021. He is currently co-Editor-in-Chief of Journal of the Franklin Institute. He acted as a Technical co-Chair for EUSIPCO 2006 and a Tutorials co-Chair of ICASSP 2014. He is a vice chair of the IEEE Technical Committees (TC) on on Image, Video and Multidimensional Signal Processing and member of TC on Machine Learning for Signal Processing, on Signal Processing Theory and Methods, and He is also a member of the IEEE Data Collections and Challenges Committee. He was a plenary speaker at ISSPA 2010, IEEE SIU 2017, Entropy 2018, MIIS 2020, IET IRC 2023 and tutorial speaker at IEEE ICSPCC 2012. He was an Alexander von Humboldt Experienced Research Fellow in the Max Planck Institute for Molecular Genetics in 2013-2015. His research interests are in the areas of statistical signal and image processing, Bayesian machine learning and information theory with applications in remote sensing, environmental sciences, telecommunications and computational biology.

\end{IEEEbiography}
\vfill

\end{document}